\documentclass[aps,prd,a4paper,preprintnumbers,tightenlines,superscriptaddress,
twocolumn,floatfix,showpacs,final,nofootinbib]{revtex4-1}
\usepackage{graphicx}
\usepackage{longtable}
\usepackage{amsmath}
\usepackage{threeparttable}
\usepackage{amssymb}
\usepackage{subfigure}
\usepackage{hyperref}
\usepackage{dcolumn}
\usepackage{bm}
\usepackage{color}

\newcommand{\mpl}{{M_{\rm {pl}}}}

\newcommand{\dd}{\, {\rm d}}

\newcommand*\colvec[3][]{\begin{pmatrix}\ifx\relax#1\relax\else#1\\\fi#2\\#3\end{pmatrix}}
\newcommand{\tg}{\tilde{g}}
\newcommand{\iii}{_{\rm i}}
\newcommand{\nm}{{\mu\nu}}
\newcommand{\eff}{_{\rm eff}}

\newcommand{\mmm}{{_{\rm m}}}

\newcommand{\dee}{{_{\rm DE}}}
\newcommand{\eee}{{_{\rm E}}}
\newcommand{\jjj}{{_{\rm J}}}

\newcommand{\tn}{\tilde{\nabla}}

\newcommand{\BBB}{B^2(\phi)}

\newcommand{\fac}{1+\frac{2\BBB X}{\Lambda^2}}
\newcommand{\facc}{1+\frac{e^{2\beta\phi}}{N^2\Lambda^2}}
\newcommand{\faccc}{1+2\frac{e^{2\beta\phi}}{N^2\Lambda^2}}

\newcommand{\pdt}{\dot{\phi}}
\newcommand{\pnm}{\partial_\mu\phi\partial_\nu\phi}
\newcommand{\den}{2 Z^2 \left(X^2 \left(3 Z^{\frac{4}{3}}-1\right)+\left(Z^{\frac{4}{3}}-1\right) \left(Z^2-Y^2\right)\right)}
\def\eea{\end{eqnarray}}
\def\bea{\begin{eqnarray}}
\allowdisplaybreaks
\usepackage{enumitem}
\begin{document}
\title{Disformal Gravity Theories: A Jordan Frame Analysis}
\author{Jeremy Sakstein}
\email[Email:]{jeremy.sakstein@port.ac.uk}
\affiliation{Institute of Cosmology and Gravitation,
University of Portsmouth, Portsmouth PO1 3FX, UK}
\author{Sarunas Verner}
\email[Email:]{sn3g11@soton.ac.uk}
\affiliation{Institute of Cosmology and Gravitation,
University of Portsmouth, Portsmouth PO1 3FX, UK}

\begin{abstract}
The Jordan frame action for general disformal theories is presented and studied for the first time, motivated by several unresolved mysteries that 
arise when working in the Einstein frame. We present the Friedmann equations and, specialising to exponential functions, study the late-time cosmology 
using dynamical systems methods and by finding approximate solutions. Our analysis reveals that either the disformal effects are irrelevant or 
the universe evolves towards a phantom phase where the equation of state of dark energy is $-3$. There is a marginal case where the asymptotic state 
of the universe depends on the model parameters and de Sitter solutions can be obtained. Our findings indicate that the metric singularity found 
using 
the Einstein frame construction corresponds phantom behaviour in the Jordan frame and we argue that this is the case for general disformal theories. 
\end{abstract}
\maketitle

\section{Introduction}
The elusive nature of dark energy has prompted a theoretical interest in the cosmological dynamics of scalar fields (see 
\cite{Copeland:2006wr} for a review) as a mechanism for driving the acceleration of the cosmic expansion. With the exception of the simplest models 
such as quintessence \cite{PhysRevD.37.3406,Zlatev:1998tr} and k-essence \cite{ArmendarizPicon:2000ah}, theories that include an additional scalar 
are alternative theories of gravity \cite{Weinberg:1965rz} (see \cite{Clifton:2011jh} for a recent compendium of cosmologically relevant theories); 
they include additional degrees of freedom that couple to matter, resulting in additional gravitational strength (or larger) interactions. 


Many (but not all) modified gravity models can be written in the schematic form
\begin{equation}
 S=\int\dd^4x\sqrt{-g}\left[\mathcal{L}_g(g_\nm)+\mathcal{L}_{\phi}(\phi)\right]+S\mmm[\tg_\nm(g_\nm,\phi)],
\end{equation}
where the scalar field $\phi$ is taken to be dimensionless. This action describes a theory of gravity in the so-called \textit{Einstein frame}. 
$\mathcal{L}_g$ contains tensor self-interactions of $g_\nm$ through curvature tensors such as the Ricci scalar and $\mathcal{L}_{\phi}$ 
contains scalar self-interactions. No direct couplings of the scalar to curvature tensors are present and the modifications of general relativity 
(GR) are encoded in the 
coupling to matter. In particular, test bodies do not move on geodesics of $\tg_\nm$, the Einstein frame metric, but instead responds to the 
composite metric $\tg_\nm(g_\nm,\phi)$---the \textit{Jordan frame} metric. It was shown by Bekenstein \cite{1992mgm..conf..905B,Bekenstein:1992pj} 
that the most general theory of a scalar coupled to matter that preserves causality is 
\begin{equation}
 \tg_\nm=C(\phi,X)g_\nm+D(\phi,X)\partial_\mu\phi\partial_\nu\phi;\quad X\equiv-\frac{1}{2}g^\nm\partial_\mu\phi\partial_\nu\phi.
\end{equation}
Indeed, it has been shown that theories where matter is coupled to metrics of this form are free of the Ostrogradski ghost instability 
\cite{Koivisto:2008ak,Zumalacarregui:2010wj,Deffayet:2009mn,Deffayet:2009wt,Deffayet:2011gz,Bettoni:2013diz,Zumalacarregui:2013pma,Gleyzes:2014qga,
Gao:2014fra,Gao:2014soa, Deffayet:2015qwa}. $C(\phi,X)$ is known as the conformal factor, and its consequences have been well-studied, at least when 
it depends on $\phi$ only. Consequently, $D(\phi,X)$ has become known as the disformal factor and the term $D\phi_\mu\phi_\nu$, the disformal 
coupling to matter, or simply the disformal part of the metric. Any theory where $D(\phi,X)\ne0$ falls into the class of disformal gravity 
theories.

Disformal couplings are ubiquitous in fundamental physics. They arise in the low energy effective action of string theory \cite{Koivisto:2013fta} and 
are linked to galileons through probe branes moving in higher dimensional space-times \cite{deRham:2010eu,Goon:2012mu}. They also arise in the 
decoupling limit of massive gravity \cite{deRham:2014zqa}. In the context of Horndeski theories \cite{Horndeski:1974wa,Deffayet:2011gz}, the most 
general scalar-tensor theories with second-order equations of motion, they are the most general transformation that preserve the form of the 
scalar-tensor sector when $C$ and $D$ depend on $\phi$ only \cite{Bettoni:2013diz}. These are the motivation behind a recent phenomenological 
study of disformal theories in several different contexts 
\cite{Kaloper:2003yf,Noller:2012sv,Zumalacarregui:2012us,vandeBruck:2012vq,Koivisto:2013fta,vandeBruck:2013yxa,Brax:2013nsa,Brax:2014vva, 
Brax:2014zba,Sakstein:2014isa,Sakstein:2014aca,Koivisto:2014gia,Hagala:2015paa,vandeBruck:2015ida,Domenech:2015hka,Brax:2015fya,Brax:2015hma, 
Tsujikawa:2015upa,Ip:2015qsa}, with most attention focusing on the case 
where $C$ and $D$ depend on $\phi$ only.

The parametrised post-Newtonian (PPN) parameters for this class of disformal theories were calculated in \cite{Sakstein:2014isa,Ip:2015qsa}, where it 
was shown that they are 
completely determined by the cosmological scalar $\phi_0$. For this reason, knowledge of the cosmology of these theories is vital for determining 
their viability. The first steps towards this were made by \cite{Sakstein:2014aca}, who used dynamical system techniques to classify 
the cosmological solution space in the Einstein frame with the goal of identifying models that passed solar system bounds. Several new fixed points 
were found but all had one problem in common: the Jordan frame metric becomes increasingly singular as the fixed point is approached, corresponding 
to the lapse in the Jordan frame approaching zero. This may represent several pathologies with the theory including the lack of a 
non-relativistic limit and a freezing of the proper time for observers, all of which we will discuss in more detail in the next section. 

Currently, it is unclear whether the singularity is debilitating for the theory or if it is merely an artifact of working in the 
Einstein frame. The two frames are equivalent for calculational purposes provided that the solutions are interpreted appropriately when relating them 
to observations. Furthermore, one cannot interpret 
observations until the proper time for an observer is aligned with the coordinate time and, since matter moves on geodesics of the Jordan frame 
metric, this frame is singled out for observations\footnote{Note that we do not claim that one frame is any more physical than the other, only that 
the Jordan frame is the frame where the coordinate time is aligned with the proper time for an observer, thus making any calculations simple to 
compare with other alternate gravity theories.} \cite{Ip:2015qsa}. Finally, Wetterich \cite{Wetterich:2014bma} has shown that frame 
transformations may introduce spurious solutions that solve the field equations in one frame but not the other. Given this, a study of the Jordan 
frame cosmology with a view to addressing these outstanding issues is certainly merited.

This is the purpose of this paper. In the next section, we introduce disformal gravity theories in the Einstein frame and discuss the nature of the 
singularity, including the associated pathologies, in more detail. The applicability of the Einstein frame dynamical system to fundamental observers 
is also discussed. We next move on to study the Jordan frame cosmology. In section \ref{sec:JF} we 
present the Jordan frame Friedmann and Klein-Gordon equations and use them to develop a dynamical systems approach to classifying the solutions for 
exponential scalar potentials and disformal factors. We 
find that theories where the disformal factor is small (in a manner to be made precise below) behave in a similar manner to quintessence but theories 
where the disformal factor is large are not well described by a dynamical systems analysis in the sense that the fixed points reveal little about the 
late-time cosmology. Instead, we focus on finding exact solutions at 
late times in section \ref{sec:ltsols}. Here, we show that the theory exhibits phantom behaviour at late times with an effective dark energy equation 
of state $w=-3$. 

Models that exhibit phantom behaviour are precisely those that suffer from singularities in the 
Einstein frame and thus we conclude that the singularity is indeed a physical pathology, the Jordan frame manifestation being phantom behaviour. 
There is a marginal case that corresponds to a specific tuning in the parameter space of the theory. In this case, the asymptotic state of the 
universe is a function of the model parameters and we show that it is possible to achieve asymptotically de Sitter solutions using a suitable tuning. 
We discuss our findings and conclude in section \ref{sec:concs}. In particular, we argue that the qualitative features we observer 
here---quintessence fixed points and phantom behaviour---are features of general disformal models.

For the reader interested purely in the cosmology of disformal models and not the singular nature of the disformal transformation, the Jordan frame 
cosmology is presented here for the first time and can be found in section \ref{sec:JF} onwards. The Friedmann equations here can be used directly 
for computing quantities such as the luminosity distance-redshift relation, which requires a transformation to coordinates appropriate for comoving 
observers if one uses the Einstein frame. Furthermore, the non-phantom regions of the parameter parameter space of exponential models is presented in 
the conclusions (section \ref{sec:concs}) where we also discuss the application to more general models.

\section{The Einstein Frame}\label{sec:EF}
In this section we present the Einstein frame action we will consider and use it to describe the singularity in the Jordan frame as well as the 
potential pathologies it presents. The action we will consider is
\begin{equation}\label{eq:EFact}
 S=\int \dd^4 x\sqrt{-g}\mpl^2\left[\frac{R(g)}{2}-\frac{1}{2}\nabla_\mu\phi\nabla^\mu\phi-V(\phi)\right]+S\mmm[\tg_\nm].
\end{equation}
The Jordan frame metric is
\begin{equation}\label{eq:jfmet}
 \tg_\nm= g_\nm+\frac{B^2(\phi)}{\Lambda^2}\partial_\mu\phi\partial_\nu\phi.
\end{equation}
Specialising to the case of a flat Friedmann-Robertson-Walker (FRW) space-time:
\begin{equation}\label{eq:EFds}
 \dd s_{\rm E}^2= -\dd t_{\rm E}^2 + a_{\rm E}^2 \dd x_{\rm E}^2,
\end{equation}
where we use subscript $\rm E$'s and $\rm J$'s to represent Einstein and Jordan frame quantities respectively, the Jordan frame 
singularity can be seen by computing the metric determinants \cite{Bekenstein:2004ne}:
\begin{equation}\label{eq:metdet}
 \frac{\sqrt{-\tg}}{\sqrt{-g}}=\sqrt{1-\Sigma};\quad\Sigma\equiv\frac{B^2(\phi)}{\Lambda^2}\left(\frac{\dd \phi}{\dd t_{\rm E}}\right)^2.
\end{equation}
When $\Sigma=1$ the Jordan frame metric is singular. Using \eqref{eq:jfmet}, the Lapse in the Jordan frame is
\begin{equation}
 N^2=1-\Sigma
\end{equation}
so that 
\begin{equation}\label{eq:tchange}
 \dd t_{\rm J}= N\dd t_{\rm E}.
\end{equation}
One can see that, cosmologically, the metric singularity corresponds to this becoming zero. 

There are several physical issues with the approach to the singularity. First, the Jordan frame space-time is
\begin{equation}\label{eq:JFt}
 \dd\tilde{s}^2= -N^2\dd t_{\rm E}^2+a(t)^2\dd x^2
\end{equation}
and so the proper time $\tau$ for physical observers is \cite{Ip:2015qsa}
\begin{equation}
 \frac{\dd \tau}{\dd t_{\rm J}}=N,
\end{equation}
and so an observers proper time is frozen at the singularity. Furthermore, since $\tg_\nm u^\mu u^\nu=-1$, the Lorentz factor is \cite{Ip:2015qsa}
\begin{equation}
 \gamma=\frac{1}{N}\left(1-\frac{v^2}{c^2}\right),
\end{equation}
where $v^i=\dd x^i/\dd t_{\rm J}$. Typically, one derives the Newtonian behaviour of the theory by looking at the limit $v/c\ll1$. In this case 
however, this is not sufficient. As demonstrated by \cite{Ip:2015qsa}, one also requires $\Sigma\ll1$ in order to have a sensible post-Newtonian 
expansion. Since the singularity corresponds precisely to $\Sigma\rightarrow1$, this behaviour is lost as the singularity is approached and there is 
no sensible Newtonian limit. The lack of such a limit was also noted by \cite{Sakstein:2014isa,Sakstein:2014aca} using Einstein frame coordinates. In 
this case, Newtonian quantities such as the total force diverge as $\Sigma\rightarrow1$. Since $N$ is also the ratio of the speed of light to that of 
tensors, the Einstein frame interpretation of this is that there are no particles that move with non-relativistic velocities in this limit. 

%

One obvious 
question is then: why not use FRW coordinates with unit lapse from the outset in the Jordan fame? In this case there is no apparent 
singularity at the level of the metric and any potential pathology must appear through the solution of the Friedmann equations. Indeed,
our choice of coordinates such that $g_\nm$ is FRW is not a choice 
of space-time since no particles follow geodesics of $g_\nm$. Applying the change of time-coordinate \eqref{eq:tchange} to 
\eqref{eq:JFt} one has
\begin{equation}
 \dd \tilde{s}^2=-\dd t_{\rm J}^2 + a(t_{\rm J})^2\dd x^2,
\end{equation}
where $a(t_{\rm J})=a(t_{\rm J}(t_{\rm E}))$. This is an FRW space-time and so one can see that the singularity found taking $g_\nm$ to be FRW is 
simply a coordinate singularity\footnote{The reader should note that this is strictly true in the context of an isotropic and homogeneous cosmology. 
Whether or not there are other physical scenarios where a non-removable Jordan frame singularity is present is unknown, although we note that, to 
date, none have been observed. All of the pathologies that arise due to the singularity discussed in this section are the result of the cosmological 
singularity and the aim of this work is to understand the physical implications of this. For this reason, we focus entirely on the cosmological 
singularity and will not attempt to address the more general question of potential singularities elsewhere. }. The 
one remaining question is that of $a(t\jjj)$. Currently, it is not known whether or not the transformation \eqref{eq:tchange} introduces any 
singularities into the spatial part of the metric. Said another way, is there some finite time $\bar{t}_{\rm J}$ such that $a(\bar{t}_{\rm J})=0$? 
This is a difficult question if one begins in the Einstein frame. Equation \eqref{eq:tchange} is highly non-linear, and one requires an exact solution 
to provide an answer. Conversely, the Jordan frame is a perfect tool because one can classify the entire cosmological solution space using dynamical 
systems or other techniques. One then has the cosmological information that can be compared to data, as well as knowledge of any pathologies. One can 
identify the Jordan frame coordinates corresponding to the singularity found using the coordinates \eqref{eq:EFds} because applying the 
transformation \eqref{eq:tchange} one finds
\begin{equation}\label{eq:NJF}
N^2= \left[1+\frac{B^2}{\Lambda^2}\left(\frac{\dd\phi}{\dd t_{\rm J}}\right)^2\right]^{-1}
\end{equation}
and so in these coordinates the singularity corresponds to $B\dd\phi/\dd t_{\rm J}/\Lambda\rightarrow\infty$. Another advantage of working in the 
Jordan frame exclusively is the following. Wetterich \cite{Wetterich:2014bma} has pointed out that spurious solutions can exist whereby a 
specific solution may be a solution of the Einstein frame equations of motion but not the Jordan frame equations. This potential problem is mitigated 
by working in the Jordan frame from the outset. 

All of the potential problems discussed above clearly motivate our study of the Jordan frame 
cosmology of disformal theories. Ultimately, we will see that when $\dd\phi/\dd t_{\rm J}\rightarrow\infty$ i.e. in the limit where the singularity 
is present in Einstein frame time, the Universe undergoes strong phantom behaviour ($w=-3$ for exponential models) and therefore the pathologies 
associated with the singularity are physical, the Jordan frame manifestation being precisely said phantom behaviour. In terms of the Einstein frame 
coordinates \eqref{eq:EFds} one can see that as the singularity is approached, the Jordan frame lapse tends increasingly towards zero and therefore, 
for comoving observers, a large number of e-folds can pass in a small amount of proper time. When viewed in this manner, the phantom behaviour is 
hardly surprising.

Before moving on to study the Jordan frame, we end this section by discussing the use of dynamical systems in both frames. The Einstein frame 
dynamical system for exponential models was studied by \cite{Zumalacarregui:2012us,Sakstein:2014aca}. In order to achieve an autonomous system of 
equations, one uses the variable $N_{\rm E}=\ln a\eee$ as a time variable and chooses appropriate variables $x_i$ that span the phase space of the 
system. Fixed points correspond to points in the phase space where
\begin{equation}
 \frac{\dd x_i}{\dd N\eee}=0\quad \forall i.
\end{equation}
It is assumed that $t\eee$ is a monotonic function of 
$N\eee$ and that $N\eee\rightarrow\infty\Rightarrow t\eee\rightarrow\infty$, both of which are necessary for global attractors of the dynamical 
system to correspond to the asymptotic state of the system (see \cite{coley2003dynamical,Alho:2015ila} for the more technical aspects of dynamical 
systems theory). Note however that $t\eee\rightarrow\infty$ does not necessarily imply that $t\jjj\rightarrow\infty$. This depends on being able to 
integrate \eqref{eq:tchange} exactly and so fixed points in the Einstein frame do not necessarily correspond to the asymptotic future 
in the Jordan frame. One case where this can be achieved trivially is the case where $N=1$ at the fixed point. In this case, disformal effects are 
absent and the theory behaves in an identical manner to quintessence. Away from this limit, there are some important and physically relevant 
quantities that cannot be calculated using the Einstein frame. One pertinent example of this is Hubble constant
\begin{equation}
 H\jjj=\frac{H\eee}{N}.
\end{equation}
In the Einstein frame, $H\eee\rightarrow0$ at the fixed points and there are also fixed points where $N\rightarrow0$ in the same limit. These are 
fixed points corresponding to the singularity. This means that the behaviour of $H\jjj$ is undetermined. If the phase space were one-dimensional, one 
could simply use L'hopital's rule to find the asymptotic value but there is no higher-dimensional analogue of this theorem. For this reason, the 
asymptotic value depends on the phase space trajectory of the specific solution as it approaches the fixed point. In this case, the Einstein frame 
dynamical system fails to achieve its goal of predicting the universal late-time behaviour since knowledge of the fixed points alone is not sufficient 
to know the asymptotic state of the Universe. This is an artifact of working in a coordinate system where the asymptotic state of dynamical system 
does not correspond to the limit of infinite proper time as seen by comoving observers\footnote{One may wonder whether it is possible to choose 
Einstein frame coordinates to avoid this problem. Such a choice of coordinates would require working in a coordinate system where $\phi(t)$ is part of 
the Einstein frame metric and would ultimately require one to mix frame variables in the equations of motion. }. Conversely, the Jordan frame 
dynamical system is perfectly able to predict all of the physically relevant quantities precisely because the coordinates are FRW with unit lapse from 
the outset.

viewed as either a coordinate 
%
\section{The Jordan Frame}\label{sec:JF}

From here on we work exclusively in the Jordan frame. For this reason, we will drop all unnecessary subscripts and tildes; it is to be 
understood that all quantities are Jordan frame quantities. The Jordan frame action is complicated compared with the simplicity of its Einstein frame 
counterpart, as are the derivation of the field equations. For this reason, we give the calculation of the Jordan frame action and the field 
equations 
in appendix \ref{sec:jfder} and present the final results here.

We begin by defining the \textit{disformal coupling},
\begin{equation}
\beta= \frac{\dd \ln B(\phi)}{d \phi}.
\end{equation}
In general, $\beta$ can be an arbitrary function of $\phi$ but, in what follows, we set $\beta$ to be constant so that
\begin{equation}\label{veq}
\qquad B(\phi)=e^{\beta \phi}.
\end{equation}
Second, the scalar potential is
\begin{equation}
 V(\phi)=m_0^2 e^{-\lambda \phi},
\end{equation}
where $\lambda$ is a constant and $m_0$ is an, a priori arbitrary, mass scale. These choices are made so that the equations exhibit a scaling 
symmetry that allows for the existence of scaling solutions and hence the dimension of the phase space is minimal \cite{Sakstein:2014aca}. With these 
newly defined constants, one can start building the disformal model. First, we need the Friedmann equations:
\begin{align}\label{3}
3H^2&=\frac{\dot{\phi}^2}{2}+Vu+8{\pi}G{\rho_m}u^{3/2}\\
\label{4}
\dot{H}&=-\frac{\dot{\phi}^2}{2}
-4 \pi G \rho_m u^{3/2}
+\frac{B^2H}{u \Lambda^2} \left(\beta \dot{\phi}^3 + \dot{\phi} \ddot{\phi} \right),
\end{align}
which are derived in Appendix \ref{sec:field_eqns}. The variable $u$ is defined for notational convenience and is given by
\begin{equation}\label{2} 
u=1+\frac{B^2\dot{\phi}^2}{\Lambda^2}.
\end{equation}
One can already see the advantage of working in the Jordan frame; the Friedmann constraint contains the disformal scale $\Lambda$ and so it is 
possible to compactify the phase space without using unphysical variables. This is in stark contrast to the Einstein frame, where the Friedmann 
constraint is identical to that of the equivalent quintessence theory and it was necessary to use advanced techniques relating to fixed points at 
infinity to determine the late-time dynamics \cite{Sakstein:2014aca}. Next, we need the the scalar equation of 
motion, which can be is expressed as:
\begin{equation}\label{5}
 \ddot{\phi}+\left[\frac{8{\pi}G{\rho_m}B^2}{\Lambda^2}\left({\ddot{\phi}+{\beta}{\dot{\phi}}^2}\right)\right]u^{3/2}
+{V_\phi u^{2}+3H{\dot{\phi}}u=\frac{{\beta}B^2{\dot{\phi}}^4}{\Lambda^2}}.
\end{equation}
Because we are working in the Jordan frame, the scalar is minimally coupled to matter and one has the usual the continuity equation:
\begin{equation}\label{6}
\dot{\rho}_{\rm m} +3H\rho\mmm = 0.
\end{equation}
Equations (\ref{3})--(\ref{6}) contain all the necessary information about the dynamics of the system. The complexity of these equations make it 
impossible to find exact analytic solutions and one method to analyse their late-time behaviour is to use a dynamical systems analysis. These methods 
are not new in cosmology. Indeed, they were applied previously to study quintessence \cite{1998PhRvD..57.4686C} and disformal theories in the Einstein 
frame \cite{Sakstein:2014aca}. Moreover, a dynamical systems analysis is a powerful tool to calculate the late-time cosmological observables. The 
unfamiliar reader can find an introduction to dynamical systems and their use in cosmology in 
\cite{Copeland:1997et,Amendola:1999er,Holden:1999hm,coley2003dynamical,Alho:2015ila,Copeland:2006wr,Sakstein:2014aca}.

Before proceeding to formulate the equations as a dynamical system, we pause to discuss the physical observables we wish to calculate using our 
subsequent analysis. Typically one is interested in the dark energy density parameter $\Omega_{\rm DE}$ and the equation of state $w$. These are 
difficult to define and several inequivalent effective variables are often found in the literature\footnote{See 
\cite{Das:2005yj,Brax:2011qs,Brax:2011bh,Brax:2012mq,Brax:2013yja,vandeBruck:2015ida} for discussions relating to effective quantities in 
 scalar-tensor theories.}. Instead of defining effective quantities, we will look for quantities whose definition and interpretation are insensitive 
to the theory of 
gravity\footnote{By which we mean they describe properties of the FRW metric and are not found by comparing the Friedmann equations to those 
resulting from the Einstein-Hilbert action.}. 
One suitable quantity is the deceleration parameter
\begin{equation}
 q=-\frac{\ddot{a}a}{\dot{a}^2},
\end{equation}
which implies that
\begin{equation}
\frac{\dot{H}}{H^2}=-\left (1+q \right )
\end{equation}
independent of the theory of gravity.
In the case of $w$CDM, one has
\begin{equation}
q=\frac{1}{2} \left(1+3w\eff \right),
\end{equation}
where $w\eff$\footnote{We use the notation $\rm eff$ to denote a single composite quantity that describes the evolution of the universe and not an 
effective equation of state for dark energy.} depends on both $w\mmm$ and $w\dee$. This motivates the definition
\begin{equation}\label{weq}
w\eff=-1-\frac{2}{3} \frac{\dot{H}}{H^2}.
\end{equation}
$q<0$, or equivalently, $w\eff<1/3$ indicates that the cosmic expansion is accelerating and so we will use these classify the nature of the 
solutions. Formally, one may define
\begin{equation}
 \Omega_{\rm DE} \equiv 1-\Omega\mmm,
\end{equation}
and we will often refer to this quantity but the reader should be aware that this is not the same quantity that is inferred from cosmic microwave 
background or luminosity 
distance measurements\footnote{By this, we mean that the values of $w$ and $\Omega\dee$ are found by fitting the data to functional forms where $w$ 
is constant, which is not necessarily the case for disformal models.}; it is merely an indication of what is driving the evolution of the universe.

\section{Formulation as a Dynamical System}\label{sec:formdyn}

\begin{figure}[bt]
 \includegraphics[width=0.5\textwidth]{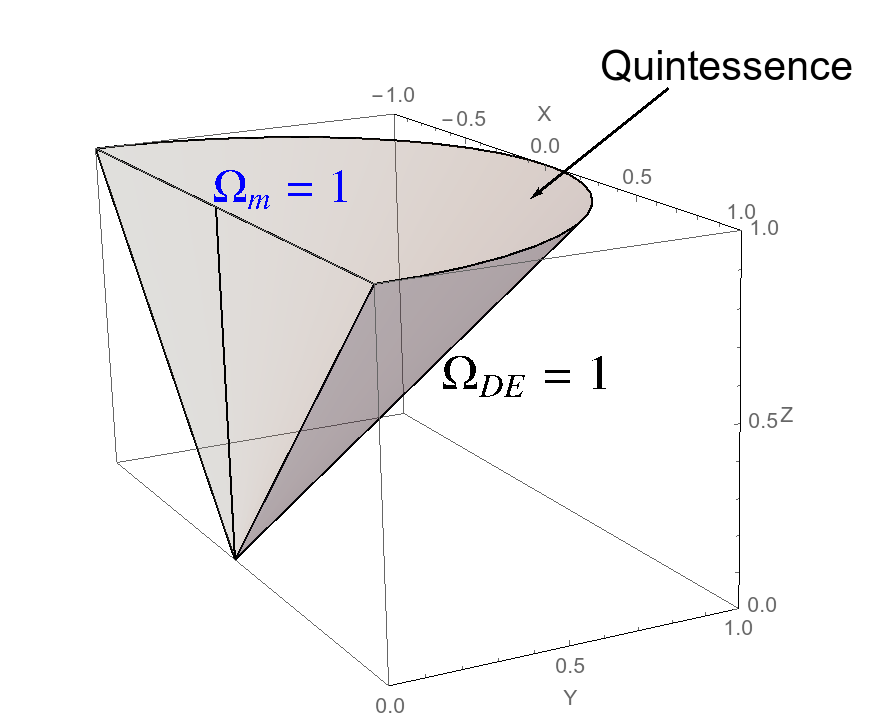}
 \caption{The phase space of the system. The point $(0,0,1)$ corresponds to matter dominated solutions and the edges of the cone correspond to dark 
energy dominated solutions. The phase space of quintessence coincides with the base of the cone located in the $Z=1$ plane.}\label{fig:phase_space}
\end{figure}

In this section, we formulate the Friedmann--Klein-Gordon equations as a dynamical system and classify the fixed points. 

\subsection{Construction of the Phase Space}

In order to make contact with the quintessence literature, 
we begin by introducing the new variables
\begin{equation}\label{7}
x\equiv \frac{\phi'}{\sqrt{6}}, \quad\textrm{and} \quad y\equiv \frac{\sqrt{V}}{\sqrt{3}H},
\end{equation}
where instead of differentiating with respect to coordinate time $t$, we differentiate with respect to $N \equiv \ln$ $a(t)$. We denote 
derivatives with respect to $N$ using a prime. This coordinate choice allows us to reduce the 
dimension of the phase space by one. Using these variables, we can rewrite the Friedmann constraint as
\begin{equation}\label{omega}
1=x^2+y^2u+\Omega_{\rm m}u^{3/2}.
\end{equation}
As noted by \cite{Sakstein:2014aca}, the disformal phase space is 
three-dimensional and so we require one more variable to close the system. \cite{Sakstein:2014aca} chose the variable $z = B H/\Lambda$ but this 
results in a phase space that is non-compact. In particular, note that $u=1+6 x^2 z^2$ in this system so that neither the $x$- nor the $z$-directions 
are compact owing to the fact that $\Omega_{\rm m}$ can be arbitrarily small. Instead, one can work in a compact phase space by introducing the 
following 
variables:
\begin{equation}\label{9} 
X\equiv xu^{-{\frac{3}{4}}}, \quad Y \equiv yu^{-\frac{1}{4}} \quad Z \equiv u^{-\frac{3}{4}}.   
\end{equation}
When written in terms of these new variables, the Friedmann constraint becomes:
\begin{equation}\label{10}
Z^2=X^2+Y^2+\Omega_m.
\end{equation}
This implies that $-1 \leq X \leq 1$, $0 \leq Y \leq 1$ and $0 \leq Z \leq 1$, therefore the phase space is compact. 
When written in terms of these variables, the phase space is a half-cone with the vertex located at $(0, 0, 0)$. This is shown in figure 
\ref{fig:phase_space}. Its base is the semicircle $x^2 + y^2 +\Omega_{\rm m}=1$ located in the plane $Z=1$. This is precisely the phase space of 
quintessence and hence corresponds to the subset of the theory where disformal effects are absent. Any fixed points that lie on the base of 
the cone therefore have late-time cosmologies that are identical to those found for pure quintessence theories with an exponential potential 
\cite{Copeland:1997et}. Note, however, that their 
stability may be altered since the three-dimensional phase space implies the existence of a third eigenvalue, and that the other two eigenvalues may 
assume different values from those found in a purely two-dimensional phase space. Setting $\Omega_{\rm m}=0$, one can see that the sides of the cone 
correspond to dark energy dominated solutions whereas setting $X=Y=0$, one has $\Omega_{\rm m} =Z=1$, and so the point $(0,0,1)$ corresponds 
to a matter dominated solution. We therefore expect all physical trajectories to originate from its vicinity.

Special attention must be paid to the tip of the cone $X=Y=Z=0$, which corresponds to what would be the metric singularity had we worked in the 
Einstein frame. This is a peculiar point because a fixed point here tells us absolutely nothing about the late-time cosmology. Typically, fixed 
points such as these indicate that the effective dimension of the phase space is reduced and one typically requires centre manifold methods to find 
the reduced phase space. Indeed, these methods are necessary for analysing the cosmology in the Einstein frame \cite{Sakstein:2014aca}. The reduced 
phase space is often unphysical\footnote{In the sense that the variables are far removed from the underlying dynamical quantities such as $H$ and 
$\dot{\phi}$. The phase space still contains all of the late-time trajectories.}, and an alternate approach is to look for approximate late-time 
solutions given that one has some idea of which terms in the equations can be ignored at late-times. This is the approach that we will adopt in 
section {\ref{sec:ltsols}. 

The case $\beta=\lambda/2$ was identified by \cite{Sakstein:2014aca} as a special parameter tuning in the Einstein frame where the dimension of the 
phase space is reduced to two. This remains the case in the Jordan frame, where one has
\begin{equation}\label{eq:reducedBV}
V(\phi)B^2(\phi)=m_0^2,
\end{equation}
which implies a relation between $X$, $Y$, and $Z$:
\begin{equation}\label{eq:consnew}
 Y^2=2\frac{m_0^2}{\Lambda^2}\frac{X^2}{1-Z^{\frac{4}{3}}}.
\end{equation}
This relation is an additional constraint that must be satisfied and hence only two of the variables are independent. In terms of the 
three-dimensional phase space, the dynamics of the system are restricted to the two-dimensional surface where \eqref{eq:consnew} is satisfied and 
hence the phase space is two-dimensional. For this reason the dynamics of this case must be treated separately.

Using equation (\ref{10}) to eliminate $\Omega\mmm$, equations (\ref{3})--(\ref{5}) can be expressed as a system of three autonomous first-order 
differential equations:
\onecolumngrid
\begin{align}
\frac{dX}{dN}&=\frac{X\left[X^4 \left(3-9 Z^{\frac{4}{3}}\right)+6 X^2 Z^{\frac{4}{3}} \left(Y^2-3 Z^{\frac{2}{3}}+4 Z^2\right)+3 
\left(Z^{\frac{4}{3}}-1\right)\left(Y^2-Z^2\right)^2\right.}{\den} \label{eq1}\\
&+\frac{\left.\sqrt{6} X Z \left(\lambda  Y^2 \left(3-5 Z^{\frac{4}{3}}\right)+2 \beta  \left(Z^{\frac{4}{3}}-1\right) (Y-Z) (Y+Z)\right)-4 \sqrt{6} 
\beta X^3 Z 
\left(Z^{\frac{4}{3}}-1\right)\right]}{\den} \notag\\
\frac{dY}{dN}&=\frac{Y\left[X^4 \left(3-9 Z^{\frac{4}{3}}\right)+6 X^2 Z^{\frac{4}{3}}\left(Y^2-3 Z^{\frac{2}{3}}+2 Z^2\right)+3 
\left(Z^{\frac{4}{3}}-1\right) 
\left(Y^2-Z^2\right)^2\right)}{\den}\label{eq2}\\
&-\frac{\left.\sqrt{6} X^3 Z \left(\lambda +6 \beta  \left(Z^{\frac{4}{3}}-1\right)-3 \lambda  Z^{\frac{4}{3}}\right)-\sqrt{6} \lambda  X 
\left(Z^{\frac{4}{3}}-1\right)Z 
\left(2 Y^2+Z^2\right)\right]}{\den}\notag\\
\label{eq3}
 \frac{dZ}{dN}&=-\frac{3 X \left(Z^{\frac{4}{3}}-1\right) \left(\sqrt{6} \left(2 \beta  X^2+\lambda  Y^2\right)-6 X Z\right)}{2 Z^4 
\left(X^2\left(1-3 
Z^{\frac{4}{3}}\right)+\left(Z^{\frac{4}{3}}-1\right) \left(Z^2-Y^2\right)\right)} .
\end{align}
Using equation (\ref{4}), one finds
\begin{align}
&\frac{H'}{H}\nonumber =\\& \frac{X^4 \left(9 Z^{\frac{4}{3}}-3\right)+4 \sqrt{6} \beta  X^3 \left(Z^{\frac{4}{3}}-1\right) Z-6 X^2 Z^{\frac{4}{3}} 
\left(Y^2-2 Z^{\frac{2}{3}}+Z^2\right)+2 \sqrt{6} \lambda  X Y^2 \left(Z^{\frac{4}{3}}-1\right) Z-3 \left(Z^{\frac{4}{3}}-1\right) 
\left(Y^2-Z^2\right)^2}{2 Z^2 \left(X^2 \left(1-3 Z^{\frac{4}{3}}\right)+\left(Z^{\frac{4}{3}}-1\right) \left(Z^2-Y^2\right)\right)},
\end{align}
\twocolumngrid
\noindent which can be used to calculate $w\eff$ and $q$. One also has $\Omega_{DE} = 1-Z^2+X^2+Y^2$ in these variables.
\allowdisplaybreaks
\subsection{Fixed Points when $\beta \neq \lambda/2$}\label{sec:fps1}

There are a total of five fixed points of equations \eqref{eq1}--\eqref{eq3} that we list in table \ref{tab:fp1}. Table \ref{tab:fpq1} lists the 
interesting cosmological quantities at each point. The corresponding eigenvalues 
are listed below; only points (4) and (5) can be late-time attractors: \newline\newline
(1) $e_1=\frac{3}{2}$, $e_2=\frac{3}{2}$, $e_3=0$\\
(2) $e_1=3$, $e_2=-2 \sqrt{6} \beta -6$, $e_3=3+\sqrt{\frac{3}{2}} \lambda$\\
(3) $e_1=3$, $e_2=2 \sqrt{6} \beta -6$, $e_3=3-\sqrt{\frac{3}{2}} \lambda$\\
(4) $e_1=\lambda  ( 2 \beta -\lambda)$, $e_2=\frac{1}{2} \left(\lambda ^2-6\right)$, $e_3=\lambda ^2-3$\\
(5) $e_1=\frac{6 \beta }{\lambda }-3$, $e_2=-\frac{3}{4}\left(1+\frac{\sqrt{24-7 \lambda ^2}}{\lambda }\right)$, \\
{\color{white}a,,..}$e_3=-\frac{3}{4}\left(1-\frac{\sqrt{24-7 \lambda ^2}}{\lambda }\right).$\\

Interestingly, not all solutions are fixed points. 
Point (1) is actually a fixed line, hence the zero eigenvalue. As 
discussed above, we will deal with this point using late-time solutions rather than dynamical systems. We note that $X=Y=Z=0$ is an independent fixed 
point not shown in the table. It corresponds to a matter dominated solution\footnote{This is point (1) in \cite{Sakstein:2014aca} table I and the 
first bullet point in \cite{Zumalacarregui:2012us}, Appendix C when the conformal parameter $\alpha=0$ (in both cases).} and so cosmologically viable 
trajectories should begin near this point. One can see that it is a saddle point and so trajectories will eventually leave its vicinity, signalling 
the onset of dark energy domination.

Points (2) and (3) are unstable nodes or saddle points that correspond to non-accelerating phases and so we will pay no further attention to them. 
Points (4) and (5) 
are both located in the plane $Z=1$, which, as discussed above, corresponds to quintessence subset. These fixed points are hence identical to the 
points found if one considers quintessence with an exponential potential in GR. In particular, point (4) is the dark-energy dominated point that 
exists when $\lambda<\sqrt{6}$. Point (5) exists when $\lambda>\sqrt{6}$ and exhibits a matter-like behaviour with $w\eff=0$. Unlike the case of GR, 
these points are not always stable when they exist. Indeed, one can see that both are unstable when $2\beta>\lambda$. When this is the case, the 
only stable fixed point is at the tip of the cone and the dynamical systems analysis does not reveal anything interesting about the 
late-time dynamics. Examples of this are shown below in figures \ref{fig:ps1} and \ref{fig:ps2}. If the theory was GR and quintessence, the models
with $\lambda=1$ and $\lambda=4$ should approach fixed points (4) and (5) independently of the other parameters. These models are plotted in figure 
\ref{fig:ps1} with $m_0=\Lambda=H_0$ and $\beta<\lambda/2$ (the model parameters are indicated in the captions). One can see that these points are 
eventually reached after a brief excursion into the domain $Z<1$. In figure \ref{fig:ps2} we plot the same models but instead choose 
$\beta>\lambda/2$. Once can see that, in this case, both models now evolve towards the tip of the cone. 

One can then conclude that models with $\beta<\lambda/2$ have late-time cosmologies that are identical to quintessence whereas those with 
$\beta>\lambda/2$ exhibit drastically different behaviour. We will calculate this below in section \ref{sec:ltsols} but we note here for completeness 
that points (2)--(5) are identical to those found by \cite{Sakstein:2014aca}\footnote{These are points (2)--(5) in that reference, table I.}. The 
reason for this is that, as discussed in section \ref{sec:EF}, these points all have $Z=1$, which corresponds to $B\dot{\phi}/\Lambda=0$ i.e. no 
disformal coupling. In this limit, the Einstein and Jordan frames are equivalent, and so are the coordinates used to parametrise the phase spaces.

\begin{table}
\bgroup
\def\arraystretch{1.5}
\begin{tabular}{|c|c|c|c|c|}\hline
    Name&X&Y&Z&Existence\\
   \hline (1)&0&0&$0 < Z \leq1$&all\\
   \hline (2)&-1&0&1&all\\
   \hline (3)&1&0&1&all\\
   \hline (4)&$\frac{\lambda}{\sqrt{6}}$&$\sqrt{1-\frac{\lambda^2}{6}}$&1&$\lambda<\sqrt{6}$\\
   \hline (5)&$\frac{\sqrt{\frac{3}{2}}}{\lambda}$&$\frac{\sqrt{\frac{3}{2}}}{\lambda}$&1&any\\
   \hline
   \end{tabular}
    \caption{The fixed points of the system \eqref{eq1}-\eqref{eq3} when $\beta \neq \lambda/2$.}\label{tab:fp1}
    \egroup
 \end{table}
 \begin{table}
  \begin{center}\bgroup
\def\arraystretch{1.5}
   \begin{tabular}{|c|c|c|c|c|}\hline
    Name&$H'/H$&$q$&$w\eff$&$\Omega_{DE}$\\
   \hline (1)&$-\frac{3}{2}$&$\frac{1}{2}$&0&1\\
   \hline (2)&-3&2&1&1\\
   \hline (3)&-3&2&1&1\\
   \hline (4)&$-\frac{\lambda ^2}{2}$&$\frac{1}{2} \left(\lambda ^2-2\right)$&$\frac{1}{3} \left(\lambda ^2-3\right)$&1\\
   \hline (5)&$-\frac{3}{2}$&$\frac{1}{2}$&0&$3/ \lambda^2$\\
   \hline
   \end{tabular}
  \caption{The cosmological variables at the fixed points of the system \eqref{eq1}-\eqref{eq3} when $\beta \neq \lambda/2$.}\label{tab:fpq1}
  \egroup
  \end{center}
 \end{table}

\begin{figure}[ht]
 \includegraphics[width=0.5\textwidth]{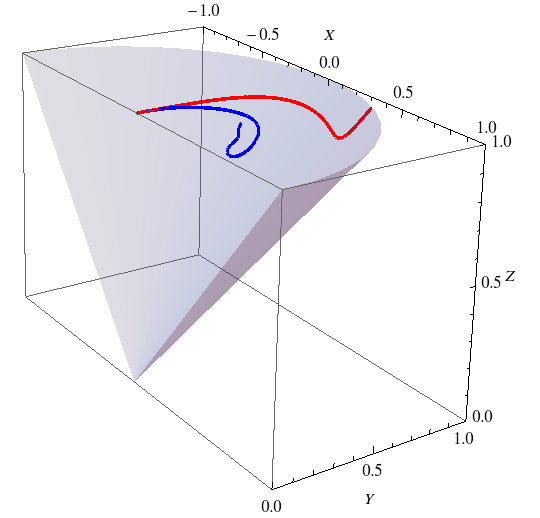}
 \caption{The phase space trajectories for models with $\lambda=1$ (red) and $\lambda = 4$ (blue). In each case $\beta=0.3$, $m_0=\Lambda=H_0$ and 
the initial conditions are $\phi(N\iii)=1$, $\phi'(N\iii)=0$. The initial value of $N\iii=\ln a\iii$ and $H(N\iii)$ were chosen such that the 
universe begins in a matter dominated phase at redshift $10$ with $\Omega\mmm=0.99999$.}\label{fig:ps1}
\end{figure}
\begin{figure}
 \includegraphics[width=0.5\textwidth]{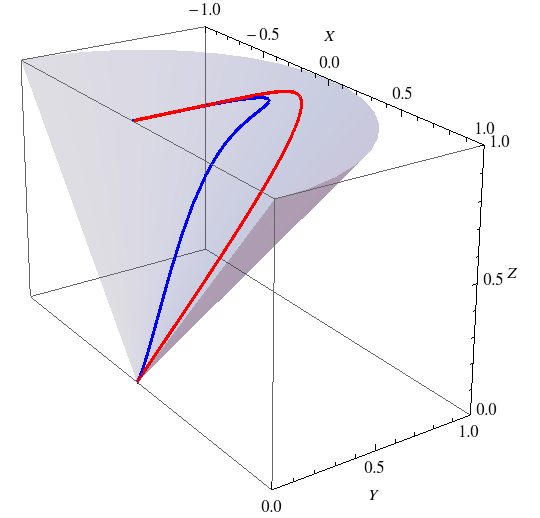}
 \caption{The phase space trajectories for models with $\lambda=1$ (red) and $\lambda = 4$ (blue). In each case $\beta=3$, $m_0=\Lambda=H_0$ and 
the initial conditions are those indicated in figure \ref{fig:ps1}.}\label{fig:ps2}
\end{figure}

\subsection{Fixed Points when $\beta=\lambda/2$}\label{sec:fpspec}

As remarked above, the phase space is two-dimensional when $\beta=\lambda/2$. To see this, we can use equation (\ref{eq:reducedBV}) in the Friedmann 
constraint \eqref{3} to find 
\begin{equation}\label{eq:2dcons}
 1=(1+2\mu^2)x^2+y^2+ \Omega\mmm\left(1+2\mu^2\frac{x^2}{y^2}\right)^{\frac{3}{2}},
\end{equation}
with $\mu=m_0/\Lambda$. In this case, the phase space is elliptical. There are two routes by which one can proceed to analyse the fixed points of the 
system. The 
first is to reformulate the equations in terms of $x$ and $y$ and use the Friedmann equation to eliminate $\Omega\mmm$. One can then find the fixed 
points of the two-dimensional system and proceed in the usual manner. The second is to continue to work in the three-dimensional framework and apply 
the constraint, which in our variables is (see equation \eqref{eq:consnew})
\begin{equation}\label{eq:cons}
 Y^2=\frac{2\mu^2X^2}{1-Z^{\frac{4}{3}}}.
\end{equation}
Here, we adopt the second approach in order to make contact with the previous analysis. Substituting the constraint \eqref{eq:cons} into equations 
\eqref{eq1}--\eqref{eq3} in order to eliminate $Y$, one finds the equations in the reduced phase space, which are given in Appendix \ref{app:reddyn} 
due to their length.

The resulting fixed points are given in table \ref{tab:fps1} with the corresponding cosmological parameters given in table \ref{tab:fpq1s}. The 
eigenvalues are
\begin{align}
{\rm  (1)}\quad e_1&=\frac{3}{2},\quad e_2=0,\\
{\rm (2)}\quad e_1&=\frac{3}{2},\quad e_2=0,\\
{\rm(3)} \quad e_1&=\frac{3}{2},\quad e_2=0,\\
{\rm(4)}\quad e_\pm&=-\frac{3}{4}\pm3\frac{\sqrt{\left(2 \left(\lambda ^4+18 
\lambda ^2-72\right) \mu ^2+72-21 \lambda ^2\right)}}{4\lambda  \sqrt{\left(2 \left(\lambda ^2-6\right) \mu ^2+3\right)}},\\
{\rm(5)}\quad e_1&=\lambda ^2-3,\quad e_2=-3+\frac{\lambda^2}{2}.
\end{align}
One can see that the first three points are unstable and so we will ignore them from here on. The fourth point is a deformation of the stable spiral 
found when $\beta<\lambda/2$ and when $\lambda>\sqrt{6}$. Its form is rather cumbersome but, by taking the limit $\lambda\rightarrow\infty$, one can 
see that the largest eigenvalue tends to zero from below and is therefore stable.\footnote{Whether or not it is an attractor or a stable spiral 
depends on the values of $\mu$ and $\lambda$.}. The fifth point is a deformation of the stable attractor found when $\beta<\lambda/2$ and when 
$\lambda<\sqrt{3}$. One can see that when $\mu>\sqrt{2}$ and $\lambda>\sqrt{3}$ the only stable point is the tip of the cone. Just like the analysis 
of the case $\beta\ne\lambda/2$, this implies that the trajectories approach a centre manifold at late times. Again, we will analyse this case by 
looking for an approximate late-time solution. The three possible types of solution are shown in figure \ref{fig:pss}. 
\begin{table*}\centering
\bgroup
\def\arraystretch{1.8}
\begin{tabular}{|c|c|c|c|c|}\hline
    Name&X&Y&Z&Existence\\
   \hline (1)&$0$&$0$&$0 < Z \leq1$& all\\
   \hline (2)&$0$&$0$&$0$&all\\
   \hline (3)&$0$&$0$&$1$&all\\
   \hline (4)&$\sqrt{\frac{3}{2\lambda^2}}{ \left(1-2 \mu ^2\right)^{3/4}}$&$\sqrt{\frac{3}{2\lambda^2}}{ \left(1-2 \mu ^2\right)^{3/4}}$&$\left(1-2 
\mu ^2\right)^{3/4}$&$\mu<\frac{1}{\sqrt{2}},\quad \lambda\ge\sqrt{3}$\\
   \hline (5)&$\frac{\lambda}{\sqrt{6}}\left(\frac{2 \lambda ^2 \mu ^2}{\lambda ^2-6}+1\right)^{3/4}$&$\sqrt{1-\frac{\lambda^2}{6}}\left(\frac{2 
\lambda ^2 \mu ^2}{\lambda ^2-6}+1\right)^{3/4}$&$\left(\frac{2 \lambda ^2 \mu ^2}{\lambda 
^2-6}+1\right)^{3/4}$&$\lambda^2<\frac{6}{1+2\mu^2}$\\
   \hline
   \end{tabular}
    \caption{The fixed points and lines when $\beta =\lambda/2$.}\label{tab:fps1}
    \egroup
 \end{table*}
 
\begin{table}
  \begin{center}
  \bgroup
\def\arraystretch{1.5}
   \begin{tabular}{|c|c|c|c|c|}\hline
    Name&$H'/H$&$q$&$w\eff$&$\Omega_{DE}$\\
   \hline (1)&$-\frac{3}{2}$&$\frac{1}{2}$&$0$&$1-Z^2$\\
   \hline (2)&$-\frac{3}{2}$&$\frac{1}{2}$&$0$&$1$\\
   \hline (3)&$-\frac{3}{2}$&$\frac{1}{2}$&$0$&$0$\\
   \hline (4)&$-\frac{3}{2}$&$\frac{1}{2}$&$ \frac{\lambda ^2}{3}-1$&$ \left(\frac{3 }{\lambda ^2}-1\right)\left(1-2 \mu ^2\right)^{3/2}+1$\\
   \hline (5)&$-\frac{\lambda ^2}{2} $&$ \frac{\lambda ^2}{2}-1$&0&$1 $\\
   \hline
   \end{tabular}
  \caption{The cosmological variables at the fixed points when $\beta = \lambda/2$.}\label{tab:fpq1s}
  \egroup
  \end{center}
 \end{table}
\begin{figure}[h]
 \includegraphics[width=0.5\textwidth]{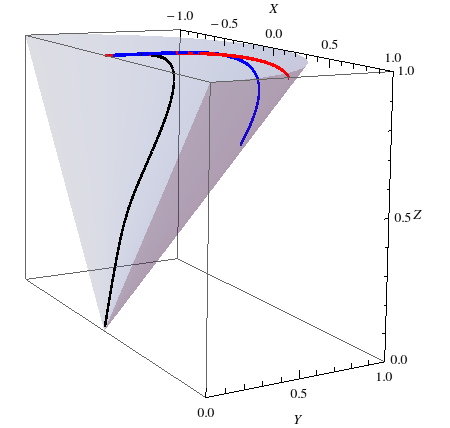}
 \caption{The phase space trajectories for possible solutions when $\beta=\lambda/2$. The blue line tends to fixed point (4) and corresponds to a 
model with $\mu=1$, $\lambda=1$. The red line tends to fixed point (5) and corresponds to a model with $\lambda=1$, $\mu=0.5$. The black line 
corresponds to a model with $\mu=1$, $\lambda=10$ and tends towards the tip of the cone. In each case $\Lambda=H_0$ and $m_0$ was fixed using the 
value of $\mu$. The initial conditions are those indicated in figure \ref{fig:ps1}.}\label{fig:pss}
\end{figure}

\section{Late-Time Solutions}\label{sec:ltsols}


In this section we address models that evolve towards the tip of the cone by looking for approximate late-time solutions. These were the cases 
$\beta>\lambda/2$ and $\beta=\lambda/2$.

\subsection{Solution when $\beta > \lambda/2$}\label{sec:5A}

At late times, one expects that the field has rolled down the potential sufficiently such that $\phi\gg1$ and $\Omega\mmm\ll 1$. Writing the 
Friedmann constraint \eqref{3} as
\begin{equation}
 3H^2=\frac{\dot{\phi}^2}{2}\left(1+\frac{m_0^2e^{(2\beta-\lambda)\phi}}{\Lambda^2}\right)+m_0^2e^{-\lambda\phi}+8\pi G \rho u^{\frac{3}{2}},
\end{equation}
one can see that the final two terms are negligible compared with the term $3H^2$ and so we have
\begin{equation}
 3H^2\approx \frac{m_0^2\pdt^2e^{(2\beta-\lambda)\phi}}{\Lambda^2}.
\end{equation}
Changing from coordinate time to $N=\ln a$ we have
\begin{equation}
 \frac{m_0^2\phi'^2e^{(2\beta-\lambda)\phi}}{\Lambda^2}=3,
\end{equation}
which is solved by
\begin{equation}\label{eq:phisol1}
 \phi(N)=\frac{2}{2\beta-\lambda}\ln\left(\sqrt{3}\frac{(2\beta-\lambda)}{2}\frac{\Lambda}{m_0}N\right).
\end{equation}
This approximate solution is shown in figure \ref{fig:ltphisol1} and one can see that it matches very closely with the numerical solution. Next, we 
can make the same approximations to equation \eqref{4} to find
\begin{equation}
 \frac{\dot{H}}{H^2}=-\frac{3}{2}\Omega\mmm\frac{\pdt^3}{\Lambda^3}e^{3\beta\phi}+\frac{1}{H}\left(\beta\pdt+\frac{\ddot{\phi}}{\pdt}\right),
\end{equation}
which, when written using $N$ as the time coordinate and applying the solution \eqref{eq:phisol1}, becomes,
\begin{equation}
 \frac{3}{2}\Omega\mmm\phi'^3\frac{H^3}{\lambda^3}e^{3\beta\phi}=\frac{5\beta-2}{2N}.
\end{equation}
Taking the logarithm of both sides, differentiating with respect to $N$ and using the relation
\begin{equation}\label{eq:omp}
 \frac{\Omega\mmm'}{\Omega\mmm}=-3-2\frac{H'}{H}
\end{equation}
we find
\begin{equation}\label{eq:hpsol}
\frac{H'}{H}=3- \frac{2(\beta+\lambda)}{(2\beta-\lambda)N}.
\end{equation}
This is plotted in figure \ref{fig:ltHsol1} and one can again see that the approximation works very well at late times. 

We can see that when $\beta>\lambda/2$ the universe will ultimately enter a phantom phase where $H'/H$ tends to $3$, although many e-folds must 
elapse before the asymptotic value is reached. That being said, it is not necessarily the case that a large change in $N$ implies 
a large amount of coordinate time has elapsed. Indeed, recalling that $H=\dd N/\dd t$, the coordinate time is
\begin{equation}
 t(N)=\int_{N\iii}^N\frac{\dd N'}{H(N')}.
\end{equation}
Since the lapse is unity, this is the proper time for comoving observers. For non-phantom solutions such as the quintessence-like trajectories found 
in section \ref{sec:fps1}, $H$ is a decreasing function of $N$ and so 
$t(N)$ is an exponentially increasing function. The phantom solutions, on the other hand, have $H(N)$ increasing exponentially and so $t(N)$ 
is a slowly evolving function at large $N$. Physically, this means that one expects a large number of e-folds in a short amount of proper time, 
and so the asymptotic phantom state is reached very quickly. This behaviour is plotted in figure \ref{fig:TN} and can be understood by considering 
the Einstein frame. As the universe 
expands, the field begins to roll and disformal effects become increasingly important. If the field does not begin to slow, the Jordan frame lapse 
approaches zero and little coordinate time evolves, despite the fact that the scale factor and field are evolving rapidly. When viewed in this 
manner, phantom behaviour is a natural consequence of the disformal coupling. 

We end this section by noting that the solution \eqref{eq:hpsol} implies that  $q\approx -4$, or, equivalently, $ w\eff\approx -3$. An equation of 
state this negative is in strong tension with observational data \cite{Sanchez:2012sg,Ade:2013ktc,Ade:2013zuv,Betoule:2014frx,Collett:2014ola} 
but it is not necessarily the case that this value is reached at the present time. Indeed, examination of figure \ref{fig:ltHsol1} reveals that, for 
the model studied there, the asymptotic value is not reached until far into the future. Whether a model predicts that the Universe is in the 
phantom phase at the present time, or that it will undergo one at some point in the future depends on the initial conditions and model parameters 
such as $m_0$ and $\Lambda$, which do not determine the asymptotic state of the Universe but do control how quickly it is reached. For example, if 
one were to tune $m_0\gg H_0$ the field will begin to roll early on in the Universe's history and one would expect phantom behaviour today. 
Conversely, tuning $m_0\ll H_0$ will result in the field being over-damped due to Hubble friction and the phantom behaviour will only ensue far 
into the future. When fitting cosmological probes of the background expansion to data, Bayesian analysis will likely favour regions of parameter 
space where the phantom phase has not yet begun and so it is likely that cosmologically viable models can be found. Such an investigation would make 
an interesting topic for future work.

\begin{figure}[h]
 \includegraphics[width=0.45\textwidth]{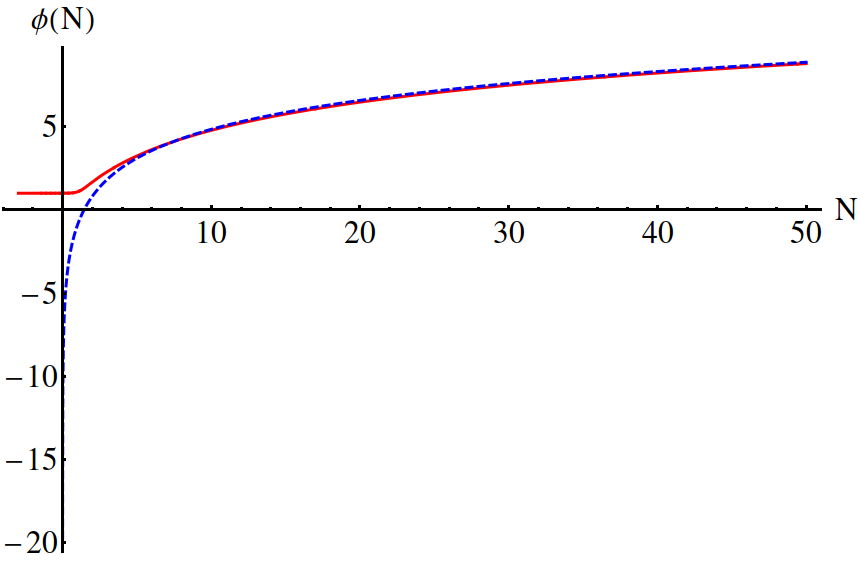}
 \caption{The evolution of $\phi(N)$ found both numerically (red solid curve) and using the approximation \eqref{eq:phisol1} (blue dashed curve). The 
parameters used were $\beta = 1.4$, $\lambda=2$ and $m_0=\Lambda=H_0$. The initial conditions are those indicated in figure 
\ref{fig:ps1}.}\label{fig:ltphisol1}
\end{figure}
\begin{figure}[h]
 \includegraphics[width=0.45\textwidth]{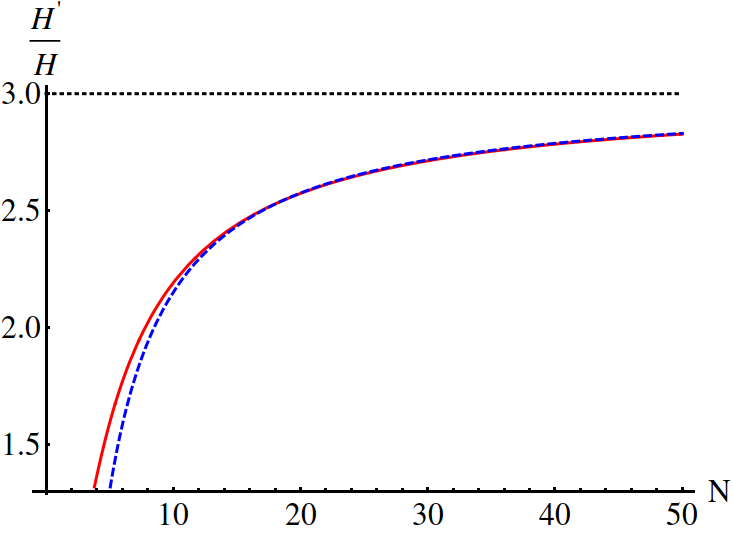}
 \caption{The evolution of $H'/H$ found both numerically (red solid curve) and using the approximation \eqref{eq:phisol1} (blue dashed curve). The 
asymptotic value of $3$ is shown using the black dotted line. The parameters used were $\beta = 1.4 $, $\lambda=2$ and $m_0=\Lambda=H_0$. The 
initial conditions are those indicated in figure \ref{fig:ps1}.}\label{fig:ltHsol1}
\end{figure}
\begin{figure}[h]
 \includegraphics[width=0.45\textwidth]{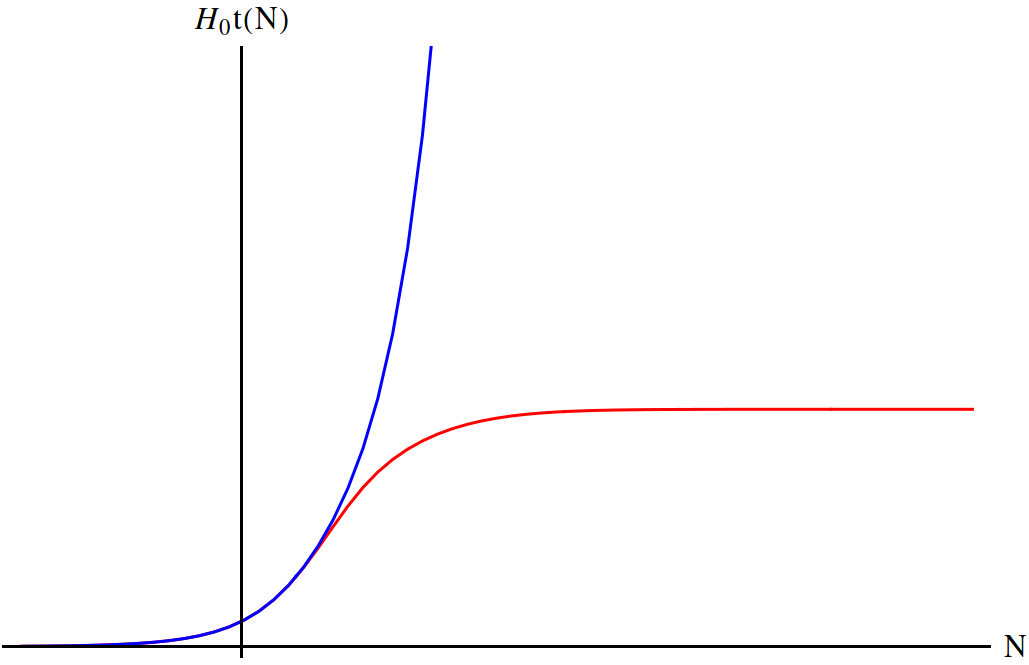}
 \caption{The coordinate time as a function of $N$ for both quintessence-like solutions (blue) and phantom solutions (red). The parameters used were 
$\beta = 1.4 $, $\lambda=2$ (red) and $\beta=0.3$, $\lambda=2$ (blue). In both cases $m_0=\Lambda=H_0$. The initial conditions are those indicated in 
figure \ref{fig:ps1}.}\label{fig:TN}
\end{figure}

\subsection{$\beta=\lambda/2$}

When $\beta=\lambda/2$ we can write the Friedmann equation as
\begin{equation}\label{eq:fried}
 3H^2=(1+2\mu^2)\frac{\pdt^2}{2}+m_0^2e^{-\lambda\phi}+8\pi G \rho u^{\frac{3}{2}}.
\end{equation}
Again, one expects that $\phi\gg1$ at late times but, unlike the previous, case, there are no factors of $e^{2\beta\phi}$ that become large in this 
limit. Instead, the second term is negligible and one has 
\begin{align}
 1&\approx(1+2\mu^2)\frac{\phi'^2}{6}+\Omega\mmm u^{\frac{3}{2}},\quad\textrm{and}\\
 0&\approx -\frac{\phi'^2}{2}-\frac{3}{2}\Omega\mmm u^{\frac{3}{2}}+\frac{\lambda}{2}\phi',
\end{align}
where the second equation comes from taking the limit $\phi\gg1$ in equation \eqref{4}. Unlike the previous case, it is not possible to find an exact 
analytic solution but one can find late-time scaling solutions by looking for 
solutions of the form $\phi'=\delta_1$, $\Omega\mmm u^{\frac{3}{2}} = \delta_2$. Under these assumptions, one is led to two equations for $\delta_i$:
\begin{align}
3&= \frac{1}{2} \left(2 \mu ^2+1\right)\delta_1^2+3 \delta_2\\
0&=\frac{\lambda  \delta_2}{2}-\frac{\delta_1^2}{2}-\frac{3 \delta_2}{2},
\end{align}
which have the solutions
\begin{align}
 (\textrm{1})\quad \delta_1&=\frac{\sqrt{\lambda ^2+12 \mu ^2-6}+\lambda }{1-2 \mu ^2}\nonumber\\\delta_2&=\frac{2}{1-2 \mu ^2}-\frac{\lambda  
\left(2 
\mu ^2+1\right) \left(\lambda -\sqrt{\lambda ^2+12 \mu ^2-6}\right)}{3 \left(1-2 \mu ^2\right)^2}\\
 (\textrm{2})\quad \delta_1&=\frac{6}{\sqrt{\lambda ^2+12 \mu ^2-6}+\lambda }\nonumber\\\delta_2&=\frac{2}{1-2 \mu ^2}-\frac{\lambda  \left(2 \mu 
^2+1\right) \left(\lambda+\sqrt{\lambda ^2+12 \mu ^2-6} \right)}{3 \left(1-2 \mu ^2\right)^2}.\label{eq:delta}
\end{align}
Note that one requires $2\mu^2>1$ in order for this type of solution to exist. If the converse is true the solution tends to fixed point (4) found 
in section \ref{sec:fpspec}. Given this constraint, one can see that solution (1) is incompatible with out assumption that $\phi\gg1$ because 
$\phi'=\delta_1<0$ and so only solution (2) is viable. A scaling relation such as this implies a definite prediction for the asymptotic state of 
the universe. Indeed, since $\Omega\mmm u^{\frac{3}{2}}$ is constant one has, using equation \eqref{eq:omp},
\begin{equation}\label{eq:HpHspec}
\frac{H'}{H} =3-\frac{3\lambda}{2}\delta_1,
\end{equation}
which implies
\begin{align}
 q&=\frac{9 \lambda }{\sqrt{\lambda ^2+12 \mu ^2-6}+\lambda }-4,\quad\textrm{and}\\
 w\eff& =\frac{6 \lambda }{\sqrt{\lambda ^2+12 \mu ^2-6}+\lambda }-3.
\end{align}
One can see that in this case the asymptotic state of the universe is a 
function of $\lambda$ and $\mu$. Note that since $\delta_1>0$, the universe cannot accelerate with $H'/H>3$. A natural question is whether 
it is possible for the universe to achieve an asymptotic de Sitter state? Setting the left hand side of \eqref{eq:HpHspec} equal to zero and using 
\eqref{eq:delta} one finds this is achieved when
\begin{equation}
 \lambda=\sqrt{2} \sqrt{2 \mu ^2-1}.
\end{equation}
As an example, we plot the evolution of $H'/H$ as a function of $N$ for the case $\mu=1$ ($\lambda=\sqrt{2}$) in figure \ref{fig:lt1hph}. One can see 
that the universe does indeed tend to a de Sitter phase at late times.

\begin{figure}[h]
 \includegraphics[width=0.45\textwidth]{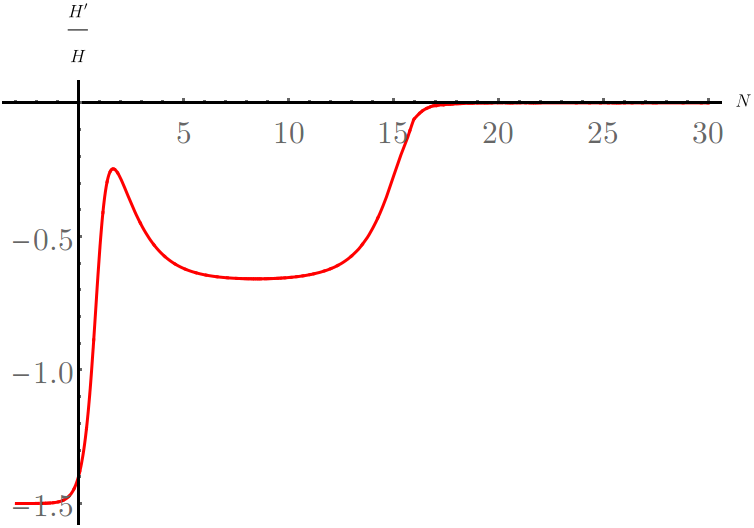}
 \caption{$H'/H$ as a function of $N$ for a model with $\mu=1$, $m_0=H_0$, $\lambda=\sqrt{2}$ and $\Lambda=H_0$. The initial conditions are those 
indicated in figure \ref{fig:ps1}.}\label{fig:lt1hph}
\end{figure}

\section{Discussion and Conclusions}\label{sec:concs}

This paper has presented and studied the Jordan frame formulation of disformal gravity theories for the first time. The Einstein frame has been 
studied extensively and motivates this study for several reasons. First, there is an apparent metric singularity that previous studies have found, 
both numerically and analytically, to be approached on cosmological scales when calculating using the Einstein frame formulation. This result has 
some pathological implications but, as discussed in section \ref{sec:EF}, it is currently unknown whether or not is it a physical pathology or merely 
an artifact of working in the Einstein frame. This paper has taken the first steps towards answering this by studying the Jordan frame cosmology and 
looking for equivalent pathologies. Second, disformal transformations from the Einstein to Jordan frame do not preserve the lapse. 
This has the result that the proper time for observers in the Jordan frame is not aligned with the coordinate time, which makes the interpretation of 
Einstein frame calculations difficult from a technical point of view. The Jordan frame does not have this problem since the lapse is unity from the 
outset.

The first part of the paper was dedicated to analysing the phase space of solutions using a dynamical systems analysis. We were successful in 
compactifying the three-dimensional phase space so that all solutions lie inside of the half-cone shown in figure \ref{fig:phase_space}. 
Interestingly, the phase space of the equivalent quintessence model (found by turning off the disformal couplings) coincides with the base of the 
cone, which allowed for transparent comparisons with quintessence. In particular, any trajectory that terminates on the base of the cone has a 
late-time cosmology that is indistinguishable from quintessence, at least at the background level. The fixed points on the base of the cone 
correspond to those found in the Einstein frame by previous studies precisely because disformal effects are absent and the time variables used to 
describe the dynamics in both frames are identical. Trajectories at the tip of the cone yield no information about the late-time cosmology 
and it was necessary to find approximate late-time solutions in order to discern the asymptotic state of the universe. In this case, one can only 
relate the Einstein and Jordan frame time variables by integrating a non-linear relation, and it is here that the power of the Jordan frame formalism 
becomes apparent. 
%

The cosmological behaviour can be summarised concisely in the $\beta$--$\lambda$ plane shown in figure \ref{fig:result}. When $\beta<\lambda/2$, all 
of the fixed points lie in the quintessence plane and so the late-time fixed points are identical to those found by \cite{Copeland:1997et}, although 
their stability is different due to the phase space being three- instead of two-dimensional. When $\beta>\lambda/2$, the only stable fixed point lies 
at the tip of the cone and so it was necessary to look for approximate late-time solutions. These were found in section \ref{sec:ltsols} where we 
showed that the universe asymptotes to a phantom state where $w\eff=-3$ ($\dot{H}/H^2=3$) independent of the model parameters. One can see from the 
various figures that the pathological 
behaviour is typically reached in the future for universes that start from matter domination and so it may be possible to reconcile the models with 
current observations. In particular, there are several model parameters, such as $m_0$ and $\Lambda$, that do not alter the position of the fixed 
points nor the stability. One would therefore expect a wide region in parameter space where the Universe is close to $\Lambda$CDM today but may 
undergo a phantom phase some time in the future. Such a model is not at odds with current observations. When fitting the model to cosmological probes 
of the background cosmology, it is likely that this region will be preferred by Bayesian fitting methods, although such analyses lie beyond the 
scope of this work. 

There is a marginal case given by $\beta=\lambda/2$ where the phase space is reduced to two. In this case we found two late-time attracting fixed 
points that lie inside the cone and one that lies at the tip. By looking for late-time scaling solutions we derived the asymptotic value of 
$\dot{H}/H^2$ for solutions that approach the tip and, in particular, were able to show that by tuning the parameters, a late-time de Sitter phase 
can be reached. 

This solution deserves further comment in light of the cosmological constant problem. In order to achieve the solution required for 
an asymptotic de Sitter phase it was necessary to tune $\beta=\lambda/2$ but this is not enough. One must further tune $\lambda$ and $\mu$ to values 
where the fixed point at the tip is the only stable one and the asymptotic value of $w\eff$ is exactly $-1$. The theory does not contain any sort of 
protective symmetry and thus the tunings required are unlikely to be technically natural. Furthermore, the model has nothing to say about the old 
cosmological constant problem because we have set all contributions to the cosmological constant from both the scalar and matter sectors to zero 
from the outset. Given this, the asymptotically de Sitter cosmological solution found here has little to say about the cosmological constant 
problem, and the fine-tuned model is hardly a compelling alternative to $\Lambda$CDM. 
  
\begin{figure}[h]
\includegraphics[width=0.45\textwidth]{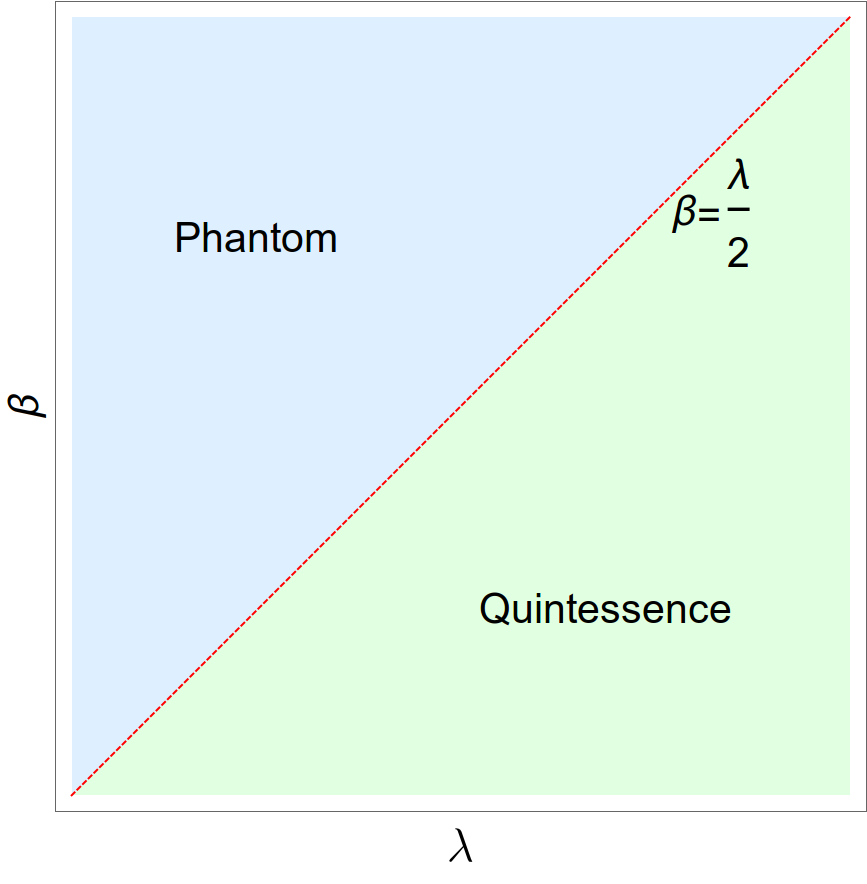}
\caption{The cosmological solutions found in this work.}\label{fig:result}
\end{figure}  
  
One of the goals of this paper is to discuss the metric singularity found by previous works using the Einstein frame formulation of the theory. The 
pathologies associated with this singularity were discussed at length in section \ref{sec:EF}. There, we noted that it is a coordinate singularity 
since one can find a gauge where the metric is perfectly regular and that it is apparently absent in the Jordan frame since one can work in this 
gauge from the outset. We showed that the singularity is located at the tip of the cone in this gauge and, furthermore, that trajectories 
approaching the tip are those that exhibit late-time phantom behaviour. The physical manifestation of the singularity is then 
clear: the universe undergoes phantom behaviour a la \cite{Caldwell:2003vq}. Retrospectively, this is somewhat to be expected from the Einstein frame 
behaviour: the approach to the singularity corresponds to the Jordan frame lapse approaching zero so that the clock for comoving observers slows 
down. A large number of e-folds can then pass in a short amount of time, which is precisely the behaviour of a phantom Universe.
%

We end by discussing the generality of our findings. In particular, the choice between a theory that is identical to quintessence, a phantom Universe 
or a finely-tuned de Sitter phase seems unappealing compared with simpler models. Here, we have only considered models where the scalar potential and 
disformal factor are exponential. This choice was made in order to yield the minimal phase space and preserve some of the scaling symmetry present in 
quintessence models. More general models will have a larger phase space that will require different variables to 
explore and one hence expects a new set of fixed points. Despite, this one would expect the qualitative features we have found here to apply. In 
particular, the fixed points were found to correspond to either phantom behaviour or the equivalent quintessence model except for a finely-tuned 
set of parameters. When written in terms of the cosmological variables and using $N=\ln a$ as a proxy for time in the Einstein frame, the disformal 
coupling leaves the spatial component unchanged but the Jordan frame lapse is given by $N^2=1-B(\phi)^2H^2\phi'^2/\Lambda^2$. All non-phantom 
Universes have $H\rightarrow0$ at late times and so one expects a set of fixed points corresponding to the equivalent quintessence models precisely 
because the disformal coupling is set zero dynamically and the Jordan and Einstein frames are equivalent. One can then discern the requirements for 
the existence of new fixed points corresponding to non-zero disformal couplings: either the Einstein frame Universe must be phantom so that $H$ or 
$\phi'$ increases without bound, or $B(\phi)$ must be chosen such that it is a strongly increasing function of $\phi$. This was the case with 
exponential models. Indeed, here we found that the disformal factor was only non zero for values of $\beta$ that were large enough to compensate for 
the decreasing of $H$. One can then see that phantom behaviour is expected for any model where $B(\phi)$ can increase rapidly enough at late times, 
what is not universal is the prediction that $w\eff=-3$, which is likely to be a theory-dependent prediction. Said another way, one can design 
models that do not exhibit phantom behaviour by construction. One simple example of this is simply $B=1$, which shows only quintessence fixed points. 
A more general example is the case of monomial potentials $V(\phi)\sim \phi^n$, $B(\phi)\sim \phi^m$ (with $n$ and $m$ positive even integers). In 
this case, one would expect $\phi$ to roll to the minimum of the potential located at $\phi=0$ at late times so that $B(\phi)$ tends to zero and the 
system behaves like quintessence. In light of the discussion above, we conclude that the general features found here---quintessence-like fixed points 
and phantom behaviour---are properties of more general disformal dark energy models.

In order to find fixed points that were neither quintessence models nor phantom Universes it was necessary to fine-tune several model parameters to 
specific values. This corresponded to reducing the dimension of the phase space so that $BH/\Lambda$ was fixed by the kinetic and potential energy. In 
this case it could neither grow without bound nor become zero. Such a fine-tuning is a very special property of the model considered here and it is 
unlikely to be a feature of more general models. More technically, the symmetries of the equations of motion ($\dot{\phi}^2\sim H^2$, $V(\phi)\sim 
H^2$ and $V_\phi\sim V(\phi)$) were crucial in allowing one to have the minimal possible phase space dimension and to identify the requisite 
parameter tunings. Constructing other theories that exhibit these features would require looking at the symmetries present when the disformal 
coupling is absent and choosing the functional form of $B(\phi)$ appropriately such that the dimension of the phase space can be preserved 
with suitable parameter tunings. It is then clear that the novel fixed points found in the marginal case are not general and require 
finely-tuned simple models to exist. 
%

We have not included a conformal factor in our analysis and it is unlikely that this will have any 
mitigating effects for the pathologies. Indeed, a conformal factor was included in the Einstein frame analysis of \cite{Sakstein:2014aca} with the 
only effect being to move the location of the fixed points. Such factors are strongly constrained by solar system tests and so the 
change is expected to be minimal. Indeed, if a conformal factor $A^2(\phi)$ is present then the Cassini constraint on the PPN parameter $\gamma$ 
\cite{Bertotti:2003rm} constrains $\dd\ln A/\dd\phi<10^{-3}$ \cite{EspositoFarese:2004cc,Ip:2015qsa}, which is the factor that appears in the 
equations governing the cosmological dynamics. 

Finally, one can relax the universal coupling and couple to dark matter only, at the cost of 
introducing violations of the equivalence principle, which are poorly constrained in the dark sector. In this case, the relevant fixed points for 
observers are those found by \cite{Sakstein:2014aca}, although the metric governing the motion of dark matter is that of a phantom universe and so one 
expects a drastic suppression of late-time structure compared with GR. 

\section*{Acknowledgements}

We would like to thank Jack Morrice, Cornelius Ramph, Fabian Schmidt, and Thomas Tram for several enlightening conversations. We are grateful to the 
SEPnet summer placement scheme for its support in organising the placement in which S.V. has participated, and to the University of Portsmouth for 
hosting S.V during the time this work was carried out.

\bibliography{ref}

\begin{thebibliography}{64}%
\makeatletter
\providecommand \@ifxundefined [1]{%
 \@ifx{#1\undefined}
}%
\providecommand \@ifnum [1]{%
 \ifnum #1\expandafter \@firstoftwo
 \else \expandafter \@secondoftwo
 \fi
}%
\providecommand \@ifx [1]{%
 \ifx #1\expandafter \@firstoftwo
 \else \expandafter \@secondoftwo
 \fi
}%
\providecommand \natexlab [1]{#1}%
\providecommand \enquote  [1]{``#1''}%
\providecommand \bibnamefont  [1]{#1}%
\providecommand \bibfnamefont [1]{#1}%
\providecommand \citenamefont [1]{#1}%
\providecommand \href@noop [0]{\@secondoftwo}%
\providecommand \href [0]{\begingroup \@sanitize@url \@href}%
\providecommand \@href[1]{\@@startlink{#1}\@@href}%
\providecommand \@@href[1]{\endgroup#1\@@endlink}%
\providecommand \@sanitize@url [0]{\catcode `\\12\catcode `\$12\catcode
  `\&12\catcode `\#12\catcode `\^12\catcode `\_12\catcode `\%12\relax}%
\providecommand \@@startlink[1]{}%
\providecommand \@@endlink[0]{}%
\providecommand \url  [0]{\begingroup\@sanitize@url \@url }%
\providecommand \@url [1]{\endgroup\@href {#1}{\urlprefix }}%
\providecommand \urlprefix  [0]{URL }%
\providecommand \Eprint [0]{\href }%
\providecommand \doibase [0]{http://dx.doi.org/}%
\providecommand \selectlanguage [0]{\@gobble}%
\providecommand \bibinfo  [0]{\@secondoftwo}%
\providecommand \bibfield  [0]{\@secondoftwo}%
\providecommand \translation [1]{[#1]}%
\providecommand \BibitemOpen [0]{}%
\providecommand \bibitemStop [0]{}%
\providecommand \bibitemNoStop [0]{.\EOS\space}%
\providecommand \EOS [0]{\spacefactor3000\relax}%
\providecommand \BibitemShut  [1]{\csname bibitem#1\endcsname}%
\let\auto@bib@innerbib\@empty
\bibitem [{\citenamefont {Copeland}\ \emph {et~al.}(2006)\citenamefont
  {Copeland}, \citenamefont {Sami},\ and\ \citenamefont
  {Tsujikawa}}]{Copeland:2006wr}%
  \BibitemOpen
  \bibfield  {author} {\bibinfo {author} {\bibfnamefont {E.~J.}\ \bibnamefont
  {Copeland}}, \bibinfo {author} {\bibfnamefont {M.}~\bibnamefont {Sami}}, \
  and\ \bibinfo {author} {\bibfnamefont {S.}~\bibnamefont {Tsujikawa}},\ }\href
  {\doibase 10.1142/S021827180600942X} {\bibfield  {journal} {\bibinfo
  {journal} {Int.J.Mod.Phys.}\ }\textbf {\bibinfo {volume} {D15}},\ \bibinfo
  {pages} {1753} (\bibinfo {year} {2006})},\ \Eprint
  {http://arxiv.org/abs/hep-th/0603057} {arXiv:hep-th/0603057 [hep-th]}
  \BibitemShut {NoStop}%
\bibitem [{\citenamefont {Ratra}\ and\ \citenamefont
  {Peebles}(1988)}]{PhysRevD.37.3406}%
  \BibitemOpen
  \bibfield  {author} {\bibinfo {author} {\bibfnamefont {B.}~\bibnamefont
  {Ratra}}\ and\ \bibinfo {author} {\bibfnamefont {P.~J.~E.}\ \bibnamefont
  {Peebles}},\ }\href {\doibase 10.1103/PhysRevD.37.3406} {\bibfield  {journal}
  {\bibinfo  {journal} {Phys. Rev. D}\ }\textbf {\bibinfo {volume} {37}},\
  \bibinfo {pages} {3406} (\bibinfo {year} {1988})}\BibitemShut {NoStop}%
\bibitem [{\citenamefont {Zlatev}\ \emph {et~al.}(1999)\citenamefont {Zlatev},
  \citenamefont {Wang},\ and\ \citenamefont {Steinhardt}}]{Zlatev:1998tr}%
  \BibitemOpen
  \bibfield  {author} {\bibinfo {author} {\bibfnamefont {I.}~\bibnamefont
  {Zlatev}}, \bibinfo {author} {\bibfnamefont {L.-M.}\ \bibnamefont {Wang}}, \
  and\ \bibinfo {author} {\bibfnamefont {P.~J.}\ \bibnamefont {Steinhardt}},\
  }\href {\doibase 10.1103/PhysRevLett.82.896} {\bibfield  {journal} {\bibinfo
  {journal} {Phys. Rev. Lett.}\ }\textbf {\bibinfo {volume} {82}},\ \bibinfo
  {pages} {896} (\bibinfo {year} {1999})},\ \Eprint
  {http://arxiv.org/abs/astro-ph/9807002} {arXiv:astro-ph/9807002 [astro-ph]}
  \BibitemShut {NoStop}%
\bibitem [{\citenamefont {Armendariz-Picon}\ \emph {et~al.}(2001)\citenamefont
  {Armendariz-Picon}, \citenamefont {Mukhanov},\ and\ \citenamefont
  {Steinhardt}}]{ArmendarizPicon:2000ah}%
  \BibitemOpen
  \bibfield  {author} {\bibinfo {author} {\bibfnamefont {C.}~\bibnamefont
  {Armendariz-Picon}}, \bibinfo {author} {\bibfnamefont {V.~F.}\ \bibnamefont
  {Mukhanov}}, \ and\ \bibinfo {author} {\bibfnamefont {P.~J.}\ \bibnamefont
  {Steinhardt}},\ }\href {\doibase 10.1103/PhysRevD.63.103510} {\bibfield
  {journal} {\bibinfo  {journal} {Phys. Rev.}\ }\textbf {\bibinfo {volume}
  {D63}},\ \bibinfo {pages} {103510} (\bibinfo {year} {2001})},\ \Eprint
  {http://arxiv.org/abs/astro-ph/0006373} {arXiv:astro-ph/0006373 [astro-ph]}
  \BibitemShut {NoStop}%
\bibitem [{\citenamefont {Weinberg}(1965)}]{Weinberg:1965rz}%
  \BibitemOpen
  \bibfield  {author} {\bibinfo {author} {\bibfnamefont {S.}~\bibnamefont
  {Weinberg}},\ }\href {\doibase 10.1103/PhysRev.138.B988} {\bibfield
  {journal} {\bibinfo  {journal} {Phys.Rev.}\ }\textbf {\bibinfo {volume}
  {138}},\ \bibinfo {pages} {B988} (\bibinfo {year} {1965})}\BibitemShut
  {NoStop}%
\bibitem [{\citenamefont {Clifton}\ \emph {et~al.}(2012)\citenamefont
  {Clifton}, \citenamefont {Ferreira}, \citenamefont {Padilla},\ and\
  \citenamefont {Skordis}}]{Clifton:2011jh}%
  \BibitemOpen
  \bibfield  {author} {\bibinfo {author} {\bibfnamefont {T.}~\bibnamefont
  {Clifton}}, \bibinfo {author} {\bibfnamefont {P.~G.}\ \bibnamefont
  {Ferreira}}, \bibinfo {author} {\bibfnamefont {A.}~\bibnamefont {Padilla}}, \
  and\ \bibinfo {author} {\bibfnamefont {C.}~\bibnamefont {Skordis}},\ }\href
  {\doibase 10.1016/j.physrep.2012.01.001} {\bibfield  {journal} {\bibinfo
  {journal} {Phys.Rept.}\ }\textbf {\bibinfo {volume} {513}},\ \bibinfo {pages}
  {1} (\bibinfo {year} {2012})},\ \Eprint {http://arxiv.org/abs/1106.2476}
  {arXiv:1106.2476 [astro-ph.CO]} \BibitemShut {NoStop}%
\bibitem [{\citenamefont {{Bekenstein}}(1992)}]{1992mgm..conf..905B}%
  \BibitemOpen
  \bibfield  {author} {\bibinfo {author} {\bibfnamefont {J.~D.}\ \bibnamefont
  {{Bekenstein}}},\ }in\ \href@noop {} {\emph {\bibinfo {booktitle} {Marcel
  Grossmann Meeting on General Relativity}}},\ \bibinfo {editor} {edited by\
  \bibinfo {editor} {\bibfnamefont {F.}~\bibnamefont {{Sat{\= o}}}}\ and\
  \bibinfo {editor} {\bibfnamefont {T.}~\bibnamefont {{Nakamura}}}}\ (\bibinfo
  {year} {1992})\ p.\ \bibinfo {pages} {905}\BibitemShut {NoStop}%
\bibitem [{\citenamefont {Bekenstein}(1993)}]{Bekenstein:1992pj}%
  \BibitemOpen
  \bibfield  {author} {\bibinfo {author} {\bibfnamefont {J.~D.}\ \bibnamefont
  {Bekenstein}},\ }\href {\doibase 10.1103/PhysRevD.48.3641} {\bibfield
  {journal} {\bibinfo  {journal} {Phys.Rev.}\ }\textbf {\bibinfo {volume}
  {D48}},\ \bibinfo {pages} {3641} (\bibinfo {year} {1993})},\ \Eprint
  {http://arxiv.org/abs/gr-qc/9211017} {arXiv:gr-qc/9211017 [gr-qc]}
  \BibitemShut {NoStop}%
\bibitem [{\citenamefont {Koivisto}(2008)}]{Koivisto:2008ak}%
  \BibitemOpen
  \bibfield  {author} {\bibinfo {author} {\bibfnamefont {T.~S.}\ \bibnamefont
  {Koivisto}},\ }\href@noop {} {\  (\bibinfo {year} {2008})},\ \Eprint
  {http://arxiv.org/abs/0811.1957} {arXiv:0811.1957 [astro-ph]} \BibitemShut
  {NoStop}%
\bibitem [{\citenamefont {Zumalacarregui}\ \emph {et~al.}(2010)\citenamefont
  {Zumalacarregui}, \citenamefont {Koivisto}, \citenamefont {Mota},\ and\
  \citenamefont {Ruiz-Lapuente}}]{Zumalacarregui:2010wj}%
  \BibitemOpen
  \bibfield  {author} {\bibinfo {author} {\bibfnamefont {M.}~\bibnamefont
  {Zumalacarregui}}, \bibinfo {author} {\bibfnamefont {T.}~\bibnamefont
  {Koivisto}}, \bibinfo {author} {\bibfnamefont {D.}~\bibnamefont {Mota}}, \
  and\ \bibinfo {author} {\bibfnamefont {P.}~\bibnamefont {Ruiz-Lapuente}},\
  }\href {\doibase 10.1088/1475-7516/2010/05/038} {\bibfield  {journal}
  {\bibinfo  {journal} {JCAP}\ }\textbf {\bibinfo {volume} {1005}},\ \bibinfo
  {pages} {038} (\bibinfo {year} {2010})},\ \Eprint
  {http://arxiv.org/abs/1004.2684} {arXiv:1004.2684 [astro-ph.CO]} \BibitemShut
  {NoStop}%
\bibitem [{\citenamefont {Deffayet}\ \emph
  {et~al.}(2009{\natexlab{a}})\citenamefont {Deffayet}, \citenamefont {Deser},\
  and\ \citenamefont {Esposito-Farese}}]{Deffayet:2009mn}%
  \BibitemOpen
  \bibfield  {author} {\bibinfo {author} {\bibfnamefont {C.}~\bibnamefont
  {Deffayet}}, \bibinfo {author} {\bibfnamefont {S.}~\bibnamefont {Deser}}, \
  and\ \bibinfo {author} {\bibfnamefont {G.}~\bibnamefont {Esposito-Farese}},\
  }\href {\doibase 10.1103/PhysRevD.80.064015} {\bibfield  {journal} {\bibinfo
  {journal} {Phys. Rev.}\ }\textbf {\bibinfo {volume} {D80}},\ \bibinfo {pages}
  {064015} (\bibinfo {year} {2009}{\natexlab{a}})},\ \Eprint
  {http://arxiv.org/abs/0906.1967} {arXiv:0906.1967 [gr-qc]} \BibitemShut
  {NoStop}%
\bibitem [{\citenamefont {Deffayet}\ \emph
  {et~al.}(2009{\natexlab{b}})\citenamefont {Deffayet}, \citenamefont
  {Esposito-Farese},\ and\ \citenamefont {Vikman}}]{Deffayet:2009wt}%
  \BibitemOpen
  \bibfield  {author} {\bibinfo {author} {\bibfnamefont {C.}~\bibnamefont
  {Deffayet}}, \bibinfo {author} {\bibfnamefont {G.}~\bibnamefont
  {Esposito-Farese}}, \ and\ \bibinfo {author} {\bibfnamefont {A.}~\bibnamefont
  {Vikman}},\ }\href {\doibase 10.1103/PhysRevD.79.084003} {\bibfield
  {journal} {\bibinfo  {journal} {Phys.Rev.}\ }\textbf {\bibinfo {volume}
  {D79}},\ \bibinfo {pages} {084003} (\bibinfo {year} {2009}{\natexlab{b}})},\
  \Eprint {http://arxiv.org/abs/0901.1314} {arXiv:0901.1314 [hep-th]}
  \BibitemShut {NoStop}%
\bibitem [{\citenamefont {Deffayet}\ \emph {et~al.}(2011)\citenamefont
  {Deffayet}, \citenamefont {Gao}, \citenamefont {Steer},\ and\ \citenamefont
  {Zahariade}}]{Deffayet:2011gz}%
  \BibitemOpen
  \bibfield  {author} {\bibinfo {author} {\bibfnamefont {C.}~\bibnamefont
  {Deffayet}}, \bibinfo {author} {\bibfnamefont {X.}~\bibnamefont {Gao}},
  \bibinfo {author} {\bibfnamefont {D.~A.}\ \bibnamefont {Steer}}, \ and\
  \bibinfo {author} {\bibfnamefont {G.}~\bibnamefont {Zahariade}},\ }\href
  {\doibase 10.1103/PhysRevD.84.064039} {\bibfield  {journal} {\bibinfo
  {journal} {Phys. Rev.}\ }\textbf {\bibinfo {volume} {D84}},\ \bibinfo {pages}
  {064039} (\bibinfo {year} {2011})},\ \Eprint {http://arxiv.org/abs/1103.3260}
  {arXiv:1103.3260 [hep-th]} \BibitemShut {NoStop}%
\bibitem [{\citenamefont {Bettoni}\ and\ \citenamefont
  {Liberati}(2013)}]{Bettoni:2013diz}%
  \BibitemOpen
  \bibfield  {author} {\bibinfo {author} {\bibfnamefont {D.}~\bibnamefont
  {Bettoni}}\ and\ \bibinfo {author} {\bibfnamefont {S.}~\bibnamefont
  {Liberati}},\ }\href {\doibase 10.1103/PhysRevD.88.084020} {\bibfield
  {journal} {\bibinfo  {journal} {Phys.Rev.}\ }\textbf {\bibinfo {volume}
  {D88}},\ \bibinfo {pages} {084020} (\bibinfo {year} {2013})},\ \Eprint
  {http://arxiv.org/abs/1306.6724} {arXiv:1306.6724 [gr-qc]} \BibitemShut
  {NoStop}%
\bibitem [{\citenamefont {Zumalacárregui}\ and\ \citenamefont
  {García-Bellido}(2014)}]{Zumalacarregui:2013pma}%
  \BibitemOpen
  \bibfield  {author} {\bibinfo {author} {\bibfnamefont {M.}~\bibnamefont
  {Zumalacárregui}}\ and\ \bibinfo {author} {\bibfnamefont {J.}~\bibnamefont
  {García-Bellido}},\ }\href {\doibase 10.1103/PhysRevD.89.064046} {\bibfield
  {journal} {\bibinfo  {journal} {Phys.Rev.}\ }\textbf {\bibinfo {volume}
  {D89}},\ \bibinfo {pages} {064046} (\bibinfo {year} {2014})},\ \Eprint
  {http://arxiv.org/abs/1308.4685} {arXiv:1308.4685 [gr-qc]} \BibitemShut
  {NoStop}%
\bibitem [{\citenamefont {Gleyzes}\ \emph {et~al.}(2014)\citenamefont
  {Gleyzes}, \citenamefont {Langlois}, \citenamefont {Piazza},\ and\
  \citenamefont {Vernizzi}}]{Gleyzes:2014qga}%
  \BibitemOpen
  \bibfield  {author} {\bibinfo {author} {\bibfnamefont {J.}~\bibnamefont
  {Gleyzes}}, \bibinfo {author} {\bibfnamefont {D.}~\bibnamefont {Langlois}},
  \bibinfo {author} {\bibfnamefont {F.}~\bibnamefont {Piazza}}, \ and\ \bibinfo
  {author} {\bibfnamefont {F.}~\bibnamefont {Vernizzi}},\ }\href@noop {} {\
  (\bibinfo {year} {2014})},\ \Eprint {http://arxiv.org/abs/1408.1952}
  {arXiv:1408.1952 [astro-ph.CO]} \BibitemShut {NoStop}%
\bibitem [{\citenamefont {Gao}(2014{\natexlab{a}})}]{Gao:2014fra}%
  \BibitemOpen
  \bibfield  {author} {\bibinfo {author} {\bibfnamefont {X.}~\bibnamefont
  {Gao}},\ }\href {\doibase 10.1103/PhysRevD.90.104033} {\bibfield  {journal}
  {\bibinfo  {journal} {Phys. Rev.}\ }\textbf {\bibinfo {volume} {D90}},\
  \bibinfo {pages} {104033} (\bibinfo {year} {2014}{\natexlab{a}})},\ \Eprint
  {http://arxiv.org/abs/1409.6708} {arXiv:1409.6708 [gr-qc]} \BibitemShut
  {NoStop}%
\bibitem [{\citenamefont {Gao}(2014{\natexlab{b}})}]{Gao:2014soa}%
  \BibitemOpen
  \bibfield  {author} {\bibinfo {author} {\bibfnamefont {X.}~\bibnamefont
  {Gao}},\ }\href {\doibase 10.1103/PhysRevD.90.081501} {\bibfield  {journal}
  {\bibinfo  {journal} {Phys. Rev.}\ }\textbf {\bibinfo {volume} {D90}},\
  \bibinfo {pages} {081501} (\bibinfo {year} {2014}{\natexlab{b}})},\ \Eprint
  {http://arxiv.org/abs/1406.0822} {arXiv:1406.0822 [gr-qc]} \BibitemShut
  {NoStop}%
\bibitem [{\citenamefont {Deffayet}\ \emph {et~al.}(2015)\citenamefont
  {Deffayet}, \citenamefont {Esposito-Farese},\ and\ \citenamefont
  {Steer}}]{Deffayet:2015qwa}%
  \BibitemOpen
  \bibfield  {author} {\bibinfo {author} {\bibfnamefont {C.}~\bibnamefont
  {Deffayet}}, \bibinfo {author} {\bibfnamefont {G.}~\bibnamefont
  {Esposito-Farese}}, \ and\ \bibinfo {author} {\bibfnamefont {D.~A.}\
  \bibnamefont {Steer}},\ }\href@noop {} {\  (\bibinfo {year} {2015})},\
  \Eprint {http://arxiv.org/abs/1506.01974} {arXiv:1506.01974 [gr-qc]}
  \BibitemShut {NoStop}%
\bibitem [{\citenamefont {Koivisto}\ \emph {et~al.}(2014)\citenamefont
  {Koivisto}, \citenamefont {Wills},\ and\ \citenamefont
  {Zavala}}]{Koivisto:2013fta}%
  \BibitemOpen
  \bibfield  {author} {\bibinfo {author} {\bibfnamefont {T.}~\bibnamefont
  {Koivisto}}, \bibinfo {author} {\bibfnamefont {D.}~\bibnamefont {Wills}}, \
  and\ \bibinfo {author} {\bibfnamefont {I.}~\bibnamefont {Zavala}},\ }\href
  {\doibase 10.1088/1475-7516/2014/06/036} {\bibfield  {journal} {\bibinfo
  {journal} {JCAP}\ }\textbf {\bibinfo {volume} {1406}},\ \bibinfo {pages}
  {036} (\bibinfo {year} {2014})},\ \Eprint {http://arxiv.org/abs/1312.2597}
  {arXiv:1312.2597 [hep-th]} \BibitemShut {NoStop}%
\bibitem [{\citenamefont {de~Rham}\ and\ \citenamefont
  {Tolley}(2010)}]{deRham:2010eu}%
  \BibitemOpen
  \bibfield  {author} {\bibinfo {author} {\bibfnamefont {C.}~\bibnamefont
  {de~Rham}}\ and\ \bibinfo {author} {\bibfnamefont {A.~J.}\ \bibnamefont
  {Tolley}},\ }\href {\doibase 10.1088/1475-7516/2010/05/015} {\bibfield
  {journal} {\bibinfo  {journal} {JCAP}\ }\textbf {\bibinfo {volume} {1005}},\
  \bibinfo {pages} {015} (\bibinfo {year} {2010})},\ \Eprint
  {http://arxiv.org/abs/1003.5917} {arXiv:1003.5917 [hep-th]} \BibitemShut
  {NoStop}%
\bibitem [{\citenamefont {Goon}\ \emph {et~al.}(2012)\citenamefont {Goon},
  \citenamefont {Hinterbichler}, \citenamefont {Joyce},\ and\ \citenamefont
  {Trodden}}]{Goon:2012mu}%
  \BibitemOpen
  \bibfield  {author} {\bibinfo {author} {\bibfnamefont {G.}~\bibnamefont
  {Goon}}, \bibinfo {author} {\bibfnamefont {K.}~\bibnamefont {Hinterbichler}},
  \bibinfo {author} {\bibfnamefont {A.}~\bibnamefont {Joyce}}, \ and\ \bibinfo
  {author} {\bibfnamefont {M.}~\bibnamefont {Trodden}},\ }\href {\doibase
  10.1016/j.physletb.2012.06.065} {\bibfield  {journal} {\bibinfo  {journal}
  {Phys. Lett.}\ }\textbf {\bibinfo {volume} {B714}},\ \bibinfo {pages} {115}
  (\bibinfo {year} {2012})},\ \Eprint {http://arxiv.org/abs/1201.0015}
  {arXiv:1201.0015 [hep-th]} \BibitemShut {NoStop}%
\bibitem [{\citenamefont {de~Rham}(2014)}]{deRham:2014zqa}%
  \BibitemOpen
  \bibfield  {author} {\bibinfo {author} {\bibfnamefont {C.}~\bibnamefont
  {de~Rham}},\ }\href@noop {} {\  (\bibinfo {year} {2014})},\ \Eprint
  {http://arxiv.org/abs/1401.4173} {arXiv:1401.4173 [hep-th]} \BibitemShut
  {NoStop}%
\bibitem [{\citenamefont {Horndeski}(1974)}]{Horndeski:1974wa}%
  \BibitemOpen
  \bibfield  {author} {\bibinfo {author} {\bibfnamefont {G.~W.}\ \bibnamefont
  {Horndeski}},\ }\href {\doibase 10.1007/BF01807638} {\bibfield  {journal}
  {\bibinfo  {journal} {Int.J.Theor.Phys.}\ }\textbf {\bibinfo {volume} {10}},\
  \bibinfo {pages} {363} (\bibinfo {year} {1974})}\BibitemShut {NoStop}%
\bibitem [{\citenamefont {Kaloper}(2004)}]{Kaloper:2003yf}%
  \BibitemOpen
  \bibfield  {author} {\bibinfo {author} {\bibfnamefont {N.}~\bibnamefont
  {Kaloper}},\ }\href {\doibase 10.1016/j.physletb.2004.01.005} {\bibfield
  {journal} {\bibinfo  {journal} {Phys.Lett.}\ }\textbf {\bibinfo {volume}
  {B583}},\ \bibinfo {pages} {1} (\bibinfo {year} {2004})},\ \Eprint
  {http://arxiv.org/abs/hep-ph/0312002} {arXiv:hep-ph/0312002 [hep-ph]}
  \BibitemShut {NoStop}%
\bibitem [{\citenamefont {Noller}(2012)}]{Noller:2012sv}%
  \BibitemOpen
  \bibfield  {author} {\bibinfo {author} {\bibfnamefont {J.}~\bibnamefont
  {Noller}},\ }\href {\doibase 10.1088/1475-7516/2012/07/013} {\bibfield
  {journal} {\bibinfo  {journal} {JCAP}\ }\textbf {\bibinfo {volume} {1207}},\
  \bibinfo {pages} {013} (\bibinfo {year} {2012})},\ \Eprint
  {http://arxiv.org/abs/1203.6639} {arXiv:1203.6639 [gr-qc]} \BibitemShut
  {NoStop}%
\bibitem [{\citenamefont {Zumalacarregui}\ \emph {et~al.}(2013)\citenamefont
  {Zumalacarregui}, \citenamefont {Koivisto},\ and\ \citenamefont
  {Mota}}]{Zumalacarregui:2012us}%
  \BibitemOpen
  \bibfield  {author} {\bibinfo {author} {\bibfnamefont {M.}~\bibnamefont
  {Zumalacarregui}}, \bibinfo {author} {\bibfnamefont {T.~S.}\ \bibnamefont
  {Koivisto}}, \ and\ \bibinfo {author} {\bibfnamefont {D.~F.}\ \bibnamefont
  {Mota}},\ }\href {\doibase 10.1103/PhysRevD.87.083010} {\bibfield  {journal}
  {\bibinfo  {journal} {Phys.Rev.}\ }\textbf {\bibinfo {volume} {D87}},\
  \bibinfo {pages} {083010} (\bibinfo {year} {2013})},\ \Eprint
  {http://arxiv.org/abs/1210.8016} {arXiv:1210.8016 [astro-ph.CO]} \BibitemShut
  {NoStop}%
\bibitem [{\citenamefont {van~de Bruck}\ and\ \citenamefont
  {Sculthorpe}(2013)}]{vandeBruck:2012vq}%
  \BibitemOpen
  \bibfield  {author} {\bibinfo {author} {\bibfnamefont {C.}~\bibnamefont
  {van~de Bruck}}\ and\ \bibinfo {author} {\bibfnamefont {G.}~\bibnamefont
  {Sculthorpe}},\ }\href {\doibase 10.1103/PhysRevD.87.044004} {\bibfield
  {journal} {\bibinfo  {journal} {Phys.Rev.}\ }\textbf {\bibinfo {volume}
  {D87}},\ \bibinfo {pages} {044004} (\bibinfo {year} {2013})},\ \Eprint
  {http://arxiv.org/abs/1210.2168} {arXiv:1210.2168 [astro-ph.CO]} \BibitemShut
  {NoStop}%
\bibitem [{\citenamefont {van~de Bruck}\ \emph {et~al.}(2013)\citenamefont
  {van~de Bruck}, \citenamefont {Morrice},\ and\ \citenamefont
  {Vu}}]{vandeBruck:2013yxa}%
  \BibitemOpen
  \bibfield  {author} {\bibinfo {author} {\bibfnamefont {C.}~\bibnamefont
  {van~de Bruck}}, \bibinfo {author} {\bibfnamefont {J.}~\bibnamefont
  {Morrice}}, \ and\ \bibinfo {author} {\bibfnamefont {S.}~\bibnamefont {Vu}},\
  }\href {\doibase 10.1103/PhysRevLett.111.161302} {\bibfield  {journal}
  {\bibinfo  {journal} {Phys.Rev.Lett.}\ }\textbf {\bibinfo {volume} {111}},\
  \bibinfo {pages} {161302} (\bibinfo {year} {2013})},\ \Eprint
  {http://arxiv.org/abs/1303.1773} {arXiv:1303.1773 [astro-ph.CO]} \BibitemShut
  {NoStop}%
\bibitem [{\citenamefont {Brax}\ \emph
  {et~al.}(2013{\natexlab{a}})\citenamefont {Brax}, \citenamefont {Burrage},
  \citenamefont {Davis},\ and\ \citenamefont {Gubitosi}}]{Brax:2013nsa}%
  \BibitemOpen
  \bibfield  {author} {\bibinfo {author} {\bibfnamefont {P.}~\bibnamefont
  {Brax}}, \bibinfo {author} {\bibfnamefont {C.}~\bibnamefont {Burrage}},
  \bibinfo {author} {\bibfnamefont {A.-C.}\ \bibnamefont {Davis}}, \ and\
  \bibinfo {author} {\bibfnamefont {G.}~\bibnamefont {Gubitosi}},\ }\href
  {\doibase 10.1088/1475-7516/2013/11/001} {\bibfield  {journal} {\bibinfo
  {journal} {JCAP}\ }\textbf {\bibinfo {volume} {1311}},\ \bibinfo {pages}
  {001} (\bibinfo {year} {2013}{\natexlab{a}})},\ \Eprint
  {http://arxiv.org/abs/1306.4168} {arXiv:1306.4168 [astro-ph.CO]} \BibitemShut
  {NoStop}%
\bibitem [{\citenamefont {Brax}\ and\ \citenamefont
  {Burrage}(2014{\natexlab{a}})}]{Brax:2014vva}%
  \BibitemOpen
  \bibfield  {author} {\bibinfo {author} {\bibfnamefont {P.}~\bibnamefont
  {Brax}}\ and\ \bibinfo {author} {\bibfnamefont {C.}~\bibnamefont {Burrage}},\
  }\href@noop {} {\  (\bibinfo {year} {2014}{\natexlab{a}})},\ \Eprint
  {http://arxiv.org/abs/1407.1861} {arXiv:1407.1861 [astro-ph.CO]} \BibitemShut
  {NoStop}%
\bibitem [{\citenamefont {Brax}\ and\ \citenamefont
  {Burrage}(2014{\natexlab{b}})}]{Brax:2014zba}%
  \BibitemOpen
  \bibfield  {author} {\bibinfo {author} {\bibfnamefont {P.}~\bibnamefont
  {Brax}}\ and\ \bibinfo {author} {\bibfnamefont {C.}~\bibnamefont {Burrage}},\
  }\href@noop {} {\  (\bibinfo {year} {2014}{\natexlab{b}})},\ \Eprint
  {http://arxiv.org/abs/1407.2376} {arXiv:1407.2376 [hep-ph]} \BibitemShut
  {NoStop}%
\bibitem [{\citenamefont {Sakstein}(2014)}]{Sakstein:2014isa}%
  \BibitemOpen
  \bibfield  {author} {\bibinfo {author} {\bibfnamefont {J.}~\bibnamefont
  {Sakstein}},\ }\href {\doibase 10.1088/1475-7516/2014/12/012} {\bibfield
  {journal} {\bibinfo  {journal} {JCAP}\ }\textbf {\bibinfo {volume} {1412}},\
  \bibinfo {pages} {012} (\bibinfo {year} {2014})},\ \Eprint
  {http://arxiv.org/abs/1409.1734} {arXiv:1409.1734 [astro-ph.CO]} \BibitemShut
  {NoStop}%
\bibitem [{\citenamefont {Sakstein}(2015)}]{Sakstein:2014aca}%
  \BibitemOpen
  \bibfield  {author} {\bibinfo {author} {\bibfnamefont {J.}~\bibnamefont
  {Sakstein}},\ }\href {\doibase 10.1103/PhysRevD.91.024036} {\bibfield
  {journal} {\bibinfo  {journal} {Phys.Rev.}\ }\textbf {\bibinfo {volume}
  {D91}},\ \bibinfo {pages} {024036} (\bibinfo {year} {2015})},\ \Eprint
  {http://arxiv.org/abs/1409.7296} {arXiv:1409.7296 [astro-ph.CO]} \BibitemShut
  {NoStop}%
\bibitem [{\citenamefont {Koivisto}\ and\ \citenamefont
  {Urban}(2014)}]{Koivisto:2014gia}%
  \BibitemOpen
  \bibfield  {author} {\bibinfo {author} {\bibfnamefont {T.~S.}\ \bibnamefont
  {Koivisto}}\ and\ \bibinfo {author} {\bibfnamefont {F.~R.}\ \bibnamefont
  {Urban}},\ }\href@noop {} {\  (\bibinfo {year} {2014})},\ \Eprint
  {http://arxiv.org/abs/1407.3445} {arXiv:1407.3445 [astro-ph.CO]} \BibitemShut
  {NoStop}%
\bibitem [{\citenamefont {Hagala}\ \emph {et~al.}(2015)\citenamefont {Hagala},
  \citenamefont {Llinares},\ and\ \citenamefont {Mota}}]{Hagala:2015paa}%
  \BibitemOpen
  \bibfield  {author} {\bibinfo {author} {\bibfnamefont {R.}~\bibnamefont
  {Hagala}}, \bibinfo {author} {\bibfnamefont {C.}~\bibnamefont {Llinares}}, \
  and\ \bibinfo {author} {\bibfnamefont {D.~F.}\ \bibnamefont {Mota}},\
  }\href@noop {} {\  (\bibinfo {year} {2015})},\ \Eprint
  {http://arxiv.org/abs/1504.07142} {arXiv:1504.07142 [astro-ph.CO]}
  \BibitemShut {NoStop}%
\bibitem [{\citenamefont {van~de Bruck}\ and\ \citenamefont
  {Morrice}(2015)}]{vandeBruck:2015ida}%
  \BibitemOpen
  \bibfield  {author} {\bibinfo {author} {\bibfnamefont {C.}~\bibnamefont
  {van~de Bruck}}\ and\ \bibinfo {author} {\bibfnamefont {J.}~\bibnamefont
  {Morrice}},\ }\href {\doibase 10.1088/1475-7516/2015/04/036} {\bibfield
  {journal} {\bibinfo  {journal} {JCAP}\ }\textbf {\bibinfo {volume} {1504}},\
  \bibinfo {pages} {036} (\bibinfo {year} {2015})},\ \Eprint
  {http://arxiv.org/abs/1501.03073} {arXiv:1501.03073 [gr-qc]} \BibitemShut
  {NoStop}%
\bibitem [{\citenamefont {Domènech}\ \emph {et~al.}(2015)\citenamefont
  {Domènech}, \citenamefont {Naruko},\ and\ \citenamefont
  {Sasaki}}]{Domenech:2015hka}%
  \BibitemOpen
  \bibfield  {author} {\bibinfo {author} {\bibfnamefont {G.}~\bibnamefont
  {Domènech}}, \bibinfo {author} {\bibfnamefont {A.}~\bibnamefont {Naruko}}, \
  and\ \bibinfo {author} {\bibfnamefont {M.}~\bibnamefont {Sasaki}},\
  }\href@noop {} {\  (\bibinfo {year} {2015})},\ \Eprint
  {http://arxiv.org/abs/1505.00174} {arXiv:1505.00174 [gr-qc]} \BibitemShut
  {NoStop}%
\bibitem [{\citenamefont {Brax}\ \emph
  {et~al.}(2015{\natexlab{a}})\citenamefont {Brax}, \citenamefont {Brun},\ and\
  \citenamefont {Wouters}}]{Brax:2015fya}%
  \BibitemOpen
  \bibfield  {author} {\bibinfo {author} {\bibfnamefont {P.}~\bibnamefont
  {Brax}}, \bibinfo {author} {\bibfnamefont {P.}~\bibnamefont {Brun}}, \ and\
  \bibinfo {author} {\bibfnamefont {D.}~\bibnamefont {Wouters}},\ }\href@noop
  {} {\  (\bibinfo {year} {2015}{\natexlab{a}})},\ \Eprint
  {http://arxiv.org/abs/1505.01020} {arXiv:1505.01020 [astro-ph.HE]}
  \BibitemShut {NoStop}%
\bibitem [{\citenamefont {Brax}\ \emph
  {et~al.}(2015{\natexlab{b}})\citenamefont {Brax}, \citenamefont {Burrage},\
  and\ \citenamefont {Englert}}]{Brax:2015hma}%
  \BibitemOpen
  \bibfield  {author} {\bibinfo {author} {\bibfnamefont {P.}~\bibnamefont
  {Brax}}, \bibinfo {author} {\bibfnamefont {C.}~\bibnamefont {Burrage}}, \
  and\ \bibinfo {author} {\bibfnamefont {C.}~\bibnamefont {Englert}},\
  }\href@noop {} {\  (\bibinfo {year} {2015}{\natexlab{b}})},\ \Eprint
  {http://arxiv.org/abs/1506.04057} {arXiv:1506.04057 [hep-ph]} \BibitemShut
  {NoStop}%
\bibitem [{\citenamefont {Tsujikawa}(2015)}]{Tsujikawa:2015upa}%
  \BibitemOpen
  \bibfield  {author} {\bibinfo {author} {\bibfnamefont {S.}~\bibnamefont
  {Tsujikawa}},\ }\href@noop {} {\  (\bibinfo {year} {2015})},\ \Eprint
  {http://arxiv.org/abs/1506.08561} {arXiv:1506.08561 [gr-qc]} \BibitemShut
  {NoStop}%
\bibitem [{\citenamefont {Ip}\ \emph {et~al.}(2015)\citenamefont {Ip},
  \citenamefont {Sakstein},\ and\ \citenamefont {Schmidt}}]{Ip:2015qsa}%
  \BibitemOpen
  \bibfield  {author} {\bibinfo {author} {\bibfnamefont {H.~Y.}\ \bibnamefont
  {Ip}}, \bibinfo {author} {\bibfnamefont {J.}~\bibnamefont {Sakstein}}, \ and\
  \bibinfo {author} {\bibfnamefont {F.}~\bibnamefont {Schmidt}},\ }\href@noop
  {} {\  (\bibinfo {year} {2015})},\ \Eprint {http://arxiv.org/abs/1507.00568}
  {arXiv:1507.00568 [gr-qc]} \BibitemShut {NoStop}%
\bibitem [{\citenamefont {Wetterich}(2014)}]{Wetterich:2014bma}%
  \BibitemOpen
  \bibfield  {author} {\bibinfo {author} {\bibfnamefont {C.}~\bibnamefont
  {Wetterich}}\ }(\bibinfo {year} {2014})\ \Eprint
  {http://arxiv.org/abs/1402.5031} {arXiv:1402.5031 [astro-ph.CO]} \BibitemShut
  {NoStop}%
\bibitem [{\citenamefont {Bekenstein}(2004)}]{Bekenstein:2004ne}%
  \BibitemOpen
  \bibfield  {author} {\bibinfo {author} {\bibfnamefont {J.~D.}\ \bibnamefont
  {Bekenstein}},\ }\href {\doibase 10.1103/PhysRevD.70.083509,
  10.1103/PhysRevD.71.069901} {\bibfield  {journal} {\bibinfo  {journal}
  {Phys.Rev.}\ }\textbf {\bibinfo {volume} {D70}},\ \bibinfo {pages} {083509}
  (\bibinfo {year} {2004})},\ \Eprint {http://arxiv.org/abs/astro-ph/0403694}
  {arXiv:astro-ph/0403694 [astro-ph]} \BibitemShut {NoStop}%
\bibitem [{\citenamefont {Coley}(2003)}]{coley2003dynamical}%
  \BibitemOpen
  \bibfield  {author} {\bibinfo {author} {\bibfnamefont {A.}~\bibnamefont
  {Coley}},\ }\href {https://books.google.co.uk/books?id=0AZCcW8Au3kC} {\emph
  {\bibinfo {title} {Dynamical Systems and Cosmology}}},\ Astrophysics and
  Space Science Library\ (\bibinfo  {publisher} {Springer Netherlands},\
  \bibinfo {year} {2003})\BibitemShut {NoStop}%
\bibitem [{\citenamefont {Alho}\ and\ \citenamefont
  {Uggla}(2015)}]{Alho:2015ila}%
  \BibitemOpen
  \bibfield  {author} {\bibinfo {author} {\bibfnamefont {A.}~\bibnamefont
  {Alho}}\ and\ \bibinfo {author} {\bibfnamefont {C.}~\bibnamefont {Uggla}},\
  }\href@noop {} {\  (\bibinfo {year} {2015})},\ \Eprint
  {http://arxiv.org/abs/1505.06903} {arXiv:1505.06903 [gr-qc]} \BibitemShut
  {NoStop}%
\bibitem [{\citenamefont {{Copeland}}\ \emph {et~al.}(1998)\citenamefont
  {{Copeland}}, \citenamefont {{Liddle}},\ and\ \citenamefont
  {{Wands}}}]{1998PhRvD..57.4686C}%
  \BibitemOpen
  \bibfield  {author} {\bibinfo {author} {\bibfnamefont {E.~J.}\ \bibnamefont
  {{Copeland}}}, \bibinfo {author} {\bibfnamefont {A.~R.}\ \bibnamefont
  {{Liddle}}}, \ and\ \bibinfo {author} {\bibfnamefont {D.}~\bibnamefont
  {{Wands}}},\ }\href {\doibase 10.1103/PhysRevD.57.4686} {\bibfield  {journal}
  {\bibinfo  {journal} {\prd}\ }\textbf {\bibinfo {volume} {57}},\ \bibinfo
  {pages} {4686} (\bibinfo {year} {1998})},\ \Eprint
  {http://arxiv.org/abs/gr-qc/9711068} {gr-qc/9711068} \BibitemShut {NoStop}%
\bibitem [{\citenamefont {Copeland}\ \emph {et~al.}(1998)\citenamefont
  {Copeland}, \citenamefont {Liddle},\ and\ \citenamefont
  {Wands}}]{Copeland:1997et}%
  \BibitemOpen
  \bibfield  {author} {\bibinfo {author} {\bibfnamefont {E.~J.}\ \bibnamefont
  {Copeland}}, \bibinfo {author} {\bibfnamefont {A.~R.}\ \bibnamefont
  {Liddle}}, \ and\ \bibinfo {author} {\bibfnamefont {D.}~\bibnamefont
  {Wands}},\ }\href {\doibase 10.1103/PhysRevD.57.4686} {\bibfield  {journal}
  {\bibinfo  {journal} {Phys.Rev.}\ }\textbf {\bibinfo {volume} {D57}},\
  \bibinfo {pages} {4686} (\bibinfo {year} {1998})},\ \Eprint
  {http://arxiv.org/abs/gr-qc/9711068} {arXiv:gr-qc/9711068 [gr-qc]}
  \BibitemShut {NoStop}%
\bibitem [{\citenamefont {Amendola}(2000)}]{Amendola:1999er}%
  \BibitemOpen
  \bibfield  {author} {\bibinfo {author} {\bibfnamefont {L.}~\bibnamefont
  {Amendola}},\ }\href {\doibase 10.1103/PhysRevD.62.043511} {\bibfield
  {journal} {\bibinfo  {journal} {Phys.Rev.}\ }\textbf {\bibinfo {volume}
  {D62}},\ \bibinfo {pages} {043511} (\bibinfo {year} {2000})},\ \Eprint
  {http://arxiv.org/abs/astro-ph/9908023} {arXiv:astro-ph/9908023 [astro-ph]}
  \BibitemShut {NoStop}%
\bibitem [{\citenamefont {Holden}\ and\ \citenamefont
  {Wands}(2000)}]{Holden:1999hm}%
  \BibitemOpen
  \bibfield  {author} {\bibinfo {author} {\bibfnamefont {D.~J.}\ \bibnamefont
  {Holden}}\ and\ \bibinfo {author} {\bibfnamefont {D.}~\bibnamefont {Wands}},\
  }\href {\doibase 10.1103/PhysRevD.61.043506} {\bibfield  {journal} {\bibinfo
  {journal} {Phys.Rev.}\ }\textbf {\bibinfo {volume} {D61}},\ \bibinfo {pages}
  {043506} (\bibinfo {year} {2000})},\ \Eprint
  {http://arxiv.org/abs/gr-qc/9908026} {arXiv:gr-qc/9908026 [gr-qc]}
  \BibitemShut {NoStop}%
\bibitem [{\citenamefont {Das}\ \emph {et~al.}(2006)\citenamefont {Das},
  \citenamefont {Corasaniti},\ and\ \citenamefont {Khoury}}]{Das:2005yj}%
  \BibitemOpen
  \bibfield  {author} {\bibinfo {author} {\bibfnamefont {S.}~\bibnamefont
  {Das}}, \bibinfo {author} {\bibfnamefont {P.~S.}\ \bibnamefont {Corasaniti}},
  \ and\ \bibinfo {author} {\bibfnamefont {J.}~\bibnamefont {Khoury}},\ }\href
  {\doibase 10.1103/PhysRevD.73.083509} {\bibfield  {journal} {\bibinfo
  {journal} {Phys. Rev.}\ }\textbf {\bibinfo {volume} {D73}},\ \bibinfo {pages}
  {083509} (\bibinfo {year} {2006})},\ \Eprint
  {http://arxiv.org/abs/astro-ph/0510628} {arXiv:astro-ph/0510628 [astro-ph]}
  \BibitemShut {NoStop}%
\bibitem [{\citenamefont {Brax}\ and\ \citenamefont
  {Davis}(2012)}]{Brax:2011qs}%
  \BibitemOpen
  \bibfield  {author} {\bibinfo {author} {\bibfnamefont {P.}~\bibnamefont
  {Brax}}\ and\ \bibinfo {author} {\bibfnamefont {A.-C.}\ \bibnamefont
  {Davis}},\ }\href@noop {} {\bibfield  {journal} {\bibinfo  {journal}
  {Phys.Lett.}\ }\textbf {\bibinfo {volume} {B707}},\ \bibinfo {pages} {1}
  (\bibinfo {year} {2012})},\ \Eprint {http://arxiv.org/abs/1109.0468}
  {arXiv:1109.0468 [hep-ph]} \BibitemShut {NoStop}%
\bibitem [{\citenamefont {Brax}\ \emph {et~al.}(2011)\citenamefont {Brax},
  \citenamefont {Davis},\ and\ \citenamefont {Winther}}]{Brax:2011bh}%
  \BibitemOpen
  \bibfield  {author} {\bibinfo {author} {\bibfnamefont {P.}~\bibnamefont
  {Brax}}, \bibinfo {author} {\bibfnamefont {A.-C.}\ \bibnamefont {Davis}}, \
  and\ \bibinfo {author} {\bibfnamefont {H.~A.}\ \bibnamefont {Winther}},\
  }\href@noop {} {\  (\bibinfo {year} {2011})},\ \Eprint
  {http://arxiv.org/abs/1112.3676} {arXiv:1112.3676 [astro-ph.CO]} \BibitemShut
  {NoStop}%
\bibitem [{\citenamefont {Brax}\ \emph
  {et~al.}(2013{\natexlab{b}})\citenamefont {Brax}, \citenamefont {Davis},\
  and\ \citenamefont {Sakstein}}]{Brax:2012mq}%
  \BibitemOpen
  \bibfield  {author} {\bibinfo {author} {\bibfnamefont {P.}~\bibnamefont
  {Brax}}, \bibinfo {author} {\bibfnamefont {A.-C.}\ \bibnamefont {Davis}}, \
  and\ \bibinfo {author} {\bibfnamefont {J.}~\bibnamefont {Sakstein}},\
  }\href@noop {} {\bibfield  {journal} {\bibinfo  {journal} {Phys. Lett. B}\
  }\textbf {\bibinfo {volume} {719}},\ \bibinfo {pages} {210} (\bibinfo {year}
  {2013}{\natexlab{b}})},\ \Eprint {http://arxiv.org/abs/1212.4392}
  {arXiv:1212.4392 [hep-th]} \BibitemShut {NoStop}%
\bibitem [{\citenamefont {Brax}\ \emph
  {et~al.}(2013{\natexlab{c}})\citenamefont {Brax}, \citenamefont {Davis},\
  and\ \citenamefont {Sakstein}}]{Brax:2013yja}%
  \BibitemOpen
  \bibfield  {author} {\bibinfo {author} {\bibfnamefont {P.}~\bibnamefont
  {Brax}}, \bibinfo {author} {\bibfnamefont {A.-C.}\ \bibnamefont {Davis}}, \
  and\ \bibinfo {author} {\bibfnamefont {J.}~\bibnamefont {Sakstein}},\
  }\href@noop {} {\  (\bibinfo {year} {2013}{\natexlab{c}})},\ \Eprint
  {http://arxiv.org/abs/1302.3080} {arXiv:1302.3080 [astro-ph.CO]} \BibitemShut
  {NoStop}%
\bibitem [{\citenamefont {Sanchez}\ \emph {et~al.}(2012)\citenamefont {Sanchez}
  \emph {et~al.}}]{Sanchez:2012sg}%
  \BibitemOpen
  \bibfield  {author} {\bibinfo {author} {\bibfnamefont {A.~G.}\ \bibnamefont
  {Sanchez}} \emph {et~al.},\ }\href {\doibase
  10.1111/j.1365-2966.2012.21502.x} {\bibfield  {journal} {\bibinfo  {journal}
  {Mon. Not. Roy. Astron. Soc.}\ }\textbf {\bibinfo {volume} {425}},\ \bibinfo
  {pages} {415} (\bibinfo {year} {2012})},\ \Eprint
  {http://arxiv.org/abs/1203.6616} {arXiv:1203.6616 [astro-ph.CO]} \BibitemShut
  {NoStop}%
\bibitem [{\citenamefont {Ade}\ \emph {et~al.}(2013{\natexlab{a}})\citenamefont
  {Ade} \emph {et~al.}}]{Ade:2013ktc}%
  \BibitemOpen
  \bibfield  {author} {\bibinfo {author} {\bibfnamefont {P.}~\bibnamefont
  {Ade}} \emph {et~al.} (\bibinfo {collaboration} {Planck Collaboration}),\
  }\href@noop {} {\  (\bibinfo {year} {2013}{\natexlab{a}})},\ \Eprint
  {http://arxiv.org/abs/1303.5062} {arXiv:1303.5062 [astro-ph.CO]} \BibitemShut
  {NoStop}%
\bibitem [{\citenamefont {Ade}\ \emph {et~al.}(2013{\natexlab{b}})\citenamefont
  {Ade} \emph {et~al.}}]{Ade:2013zuv}%
  \BibitemOpen
  \bibfield  {author} {\bibinfo {author} {\bibfnamefont {P.}~\bibnamefont
  {Ade}} \emph {et~al.} (\bibinfo {collaboration} {Planck Collaboration}),\
  }\href@noop {} {\  (\bibinfo {year} {2013}{\natexlab{b}})},\ \Eprint
  {http://arxiv.org/abs/1303.5076} {arXiv:1303.5076 [astro-ph.CO]} \BibitemShut
  {NoStop}%
\bibitem [{\citenamefont {Betoule}\ \emph {et~al.}(2014)\citenamefont {Betoule}
  \emph {et~al.}}]{Betoule:2014frx}%
  \BibitemOpen
  \bibfield  {author} {\bibinfo {author} {\bibfnamefont {M.}~\bibnamefont
  {Betoule}} \emph {et~al.} (\bibinfo {collaboration} {SDSS}),\ }\href
  {\doibase 10.1051/0004-6361/201423413} {\bibfield  {journal} {\bibinfo
  {journal} {Astron. Astrophys.}\ }\textbf {\bibinfo {volume} {568}},\ \bibinfo
  {pages} {A22} (\bibinfo {year} {2014})},\ \Eprint
  {http://arxiv.org/abs/1401.4064} {arXiv:1401.4064 [astro-ph.CO]} \BibitemShut
  {NoStop}%
\bibitem [{\citenamefont {Collett}\ and\ \citenamefont
  {Auger}(2014)}]{Collett:2014ola}%
  \BibitemOpen
  \bibfield  {author} {\bibinfo {author} {\bibfnamefont {T.~E.}\ \bibnamefont
  {Collett}}\ and\ \bibinfo {author} {\bibfnamefont {M.~W.}\ \bibnamefont
  {Auger}},\ }\href {\doibase 10.1093/mnras/stu1190} {\bibfield  {journal}
  {\bibinfo  {journal} {Mon. Not. Roy. Astron. Soc.}\ }\textbf {\bibinfo
  {volume} {443}},\ \bibinfo {pages} {969} (\bibinfo {year} {2014})},\ \Eprint
  {http://arxiv.org/abs/1403.5278} {arXiv:1403.5278 [astro-ph.CO]} \BibitemShut
  {NoStop}%
\bibitem [{\citenamefont {Caldwell}\ \emph {et~al.}(2003)\citenamefont
  {Caldwell}, \citenamefont {Kamionkowski},\ and\ \citenamefont
  {Weinberg}}]{Caldwell:2003vq}%
  \BibitemOpen
  \bibfield  {author} {\bibinfo {author} {\bibfnamefont {R.~R.}\ \bibnamefont
  {Caldwell}}, \bibinfo {author} {\bibfnamefont {M.}~\bibnamefont
  {Kamionkowski}}, \ and\ \bibinfo {author} {\bibfnamefont {N.~N.}\
  \bibnamefont {Weinberg}},\ }\href {\doibase 10.1103/PhysRevLett.91.071301}
  {\bibfield  {journal} {\bibinfo  {journal} {Phys. Rev. Lett.}\ }\textbf
  {\bibinfo {volume} {91}},\ \bibinfo {pages} {071301} (\bibinfo {year}
  {2003})},\ \Eprint {http://arxiv.org/abs/astro-ph/0302506}
  {arXiv:astro-ph/0302506 [astro-ph]} \BibitemShut {NoStop}%
\bibitem [{\citenamefont {Bertotti}\ \emph {et~al.}(2003)\citenamefont
  {Bertotti}, \citenamefont {Iess},\ and\ \citenamefont
  {Tortora}}]{Bertotti:2003rm}%
  \BibitemOpen
  \bibfield  {author} {\bibinfo {author} {\bibfnamefont {B.}~\bibnamefont
  {Bertotti}}, \bibinfo {author} {\bibfnamefont {L.}~\bibnamefont {Iess}}, \
  and\ \bibinfo {author} {\bibfnamefont {P.}~\bibnamefont {Tortora}},\ }\href
  {\doibase 10.1038/nature01997} {\bibfield  {journal} {\bibinfo  {journal}
  {Nature}\ }\textbf {\bibinfo {volume} {425}},\ \bibinfo {pages} {374}
  (\bibinfo {year} {2003})}\BibitemShut {NoStop}%
\bibitem [{\citenamefont {Esposito-Farese}(2004)}]{EspositoFarese:2004cc}%
  \BibitemOpen
  \bibfield  {author} {\bibinfo {author} {\bibfnamefont {G.}~\bibnamefont
  {Esposito-Farese}},\ }\href {\doibase 10.1063/1.1835173} {\bibfield
  {journal} {\bibinfo  {journal} {AIP Conf.Proc.}\ }\textbf {\bibinfo {volume}
  {736}},\ \bibinfo {pages} {35} (\bibinfo {year} {2004})},\ \Eprint
  {http://arxiv.org/abs/gr-qc/0409081} {arXiv:gr-qc/0409081 [gr-qc]}
  \BibitemShut {NoStop}%
\bibitem [{\citenamefont {Renaux-Petel}\ \emph {et~al.}(2011)\citenamefont
  {Renaux-Petel}, \citenamefont {Mizuno},\ and\ \citenamefont
  {Koyama}}]{RenauxPetel:2011uk}%
  \BibitemOpen
  \bibfield  {author} {\bibinfo {author} {\bibfnamefont {S.}~\bibnamefont
  {Renaux-Petel}}, \bibinfo {author} {\bibfnamefont {S.}~\bibnamefont
  {Mizuno}}, \ and\ \bibinfo {author} {\bibfnamefont {K.}~\bibnamefont
  {Koyama}},\ }\href {\doibase 10.1088/1475-7516/2011/11/042} {\bibfield
  {journal} {\bibinfo  {journal} {JCAP}\ }\textbf {\bibinfo {volume} {1111}},\
  \bibinfo {pages} {042} (\bibinfo {year} {2011})},\ \Eprint
  {http://arxiv.org/abs/1108.0305} {arXiv:1108.0305 [astro-ph.CO]} \BibitemShut
  {NoStop}%
\end{thebibliography}%

\appendix
\section{Transformation to the Jordan Frame}\label{sec:jfder}

In this appendix we transform the Einstein frame action to the Jordan frame and derive the Friedmann and Klein-Gordon equations.

\subsection{The Jordan Frame Action}

Our starting point is the Einstein frame action (\ref{eq:EFact}), which we write as
\begin{align}
 S&=\int\dd^4x{\mpl^2}\left[\mathcal{L}_{g}+\mathcal{L}_{\phi}\right]+S\mmm[\tg]\quad\textrm{with}\nonumber\\
 \mathcal{L}_{\rm g}&=\frac{\sqrt{-g}R(g)}{2}\quad\textrm{and}\nonumber\\
 \mathcal{L}_\phi&=\sqrt{-g}\left[-\frac{1}{2}\tn_\mu\phi\tn^\mu\phi-V(\phi)\right].\label{eq:efactapp}
\end{align}
Bettoni and Liberati \cite{Bettoni:2013diz} have shown that the Horndeski action \cite{Horndeski:1974wa}---the most general scalar-tensor theory with 
manifestly second-order field equations---is invariant under disformal transformations and furthermore that the Einstein frame exists only when terms 
quintic in the scalar are absent. For. this reason, we expect that the Jordan frame action takes the form
\begin{align} 
S&=\int\dd^4x\sqrt{-\tg}\left(G_2(\phi,X)+G_3(\phi,X)\Box\phi+G_4(\phi,X)R(\tg)\right.\nonumber\\&\left.+G_{4,X}\left[
(\Box\phi)^2-\tn_\mu\tn_\nu\phi\tn^\mu\tn^\nu\phi\right] \right)+S\mmm[\tg_\nm],\label{eq:horn}
\end{align}
where $G_i$ are arbitrary functions, $X=-\tg^\nm\pnm/2$, and $\Box=\tg^\nm\nabla_\mu\nabla_\nu$. Our strategy is then to transform each term in 
(\ref{eq:efactapp}) into the Jordan frame 
by inverting (\ref{eq:jfmet}) and then performing manipulations to get it into the form (\ref{eq:horn}). To accomplish this, we follow the methods of 
\cite{Zumalacarregui:2012us,Zumalacarregui:2013pma}. We begin by inverting (\ref{eq:jfmet}) to find
\begin{align}
 g_\nm&=\tg_\nm-\frac{\BBB}{\Lambda^2}\pnm\quad\textrm{and}\label{eq:tg1}\\
g^\nm&=\tg^\nm+\frac{\BBB}{\Lambda^2}\frac{\tn^\mu\phi\tn^\nu\phi}{1+\frac{2\BBB X}{\Lambda^2}},\label{eq:tg2}
\end{align}
where all contractions are performed using $\tg_\nm$. Next, we introduce the tensor
\begin{align} 
\mathcal{K}_\nm^\alpha&=\Gamma_\nm^\alpha-\tilde{\Gamma}_\nm^\alpha\\&=\frac{\BBB\tn^\alpha\phi\tilde{\nabla}_\mu\tilde{\nabla}
_\nu\phi+B(\phi)B_\phi (\phi)\tn^\alpha\phi\pnm} {\Lambda^2\left(1+\frac{2\BBB X}{\Lambda^2}\right) } .
\end{align}
using the identity $2\tilde{\nabla}_{[\mu}\tilde{\nabla}_{\nu]}v^\beta=R^\alpha_{\beta\mu\nu}v^\beta$ one finds \cite{Zumalacarregui:2013pma}
\begin{equation} 
R^\alpha_{\,\,\,\,\beta\mu\nu}=\tilde{R}^\alpha_{\,\,\,\,\beta\mu\nu}+2\tilde{\nabla}_{[\mu}\mathcal{K}^\alpha_{\,\,\,\,\nu]\beta}+2\mathcal{K}
^\alpha_{\,\,\, \,\sigma [\mu} \mathcal{K}^\sigma_{\,\,\,\,\nu]\beta},
\end{equation}
which, after making the appropriate contractions and using equations (\ref{eq:metdet}) and (\ref{eq:tg1}), can be used to transform 
$\mathcal{L}_g$:
\onecolumngrid
\begin{align}
\frac{\mathcal{L}_g}{\sqrt{-\tg}}&=\frac{\sqrt{-g}}{\sqrt{-\tg}}\left[g^\nm\left(\tilde{R}^\alpha_{\,\,\,\,\mu\alpha\nu}
-2\mathcal{K}^\alpha_{\,\,\,\,\sigma[\alpha}\mathcal{K}^\sigma_{\,\,\,\,\mu]\nu} 
\right)\right]=\sqrt{\fac}\frac{\tilde{R}}{2}+\frac{\BBB}{\Lambda^2\sqrt{1+\frac{2\BBB 
X}{\Lambda^2}}}\tilde{R}_\nm\tn^\mu\phi\tn^\nu\phi\nonumber\\&-\frac{\BBB}{2\Lambda^2\sqrt{\fac}}\left[
(\Box\phi)^2-\tn_\mu\tn_\nu\phi\tn^\mu\tn^\nu\phi\right]+\frac{B(\phi) 
B_\phi(\phi)}{\Lambda^2\left(\fac\right)^{\frac{3}{2}}}\left[\tn^\mu\phi\tilde{\nabla}_\mu\tn_\nu\phi\tn^\nu\phi+2X\Box\phi\right]\nonumber\\&+\frac{
B^4(\phi)}{\Lambda^2\left(\fac\right)^{\frac{3}{2}}}\left[
\Box\phi\tn^\mu\phi\tilde{\nabla}_\mu\tn_\nu\phi\tn^\nu\phi+\tn^\mu\phi\tn^\nu\phi\tn_\alpha\tn_\nu\phi\tn^\alpha\tn_\mu\phi 
\right].
\end{align}
\twocolumngrid
This is not yet in Horndeski form; there are two quartic and one cubic term that needs to be removed. Furthermore, there is a term 
proportional to 
$\tilde{R}_\nm\tn^\mu\phi\tn^\nu\phi$. We can remove this term and the quartic one by adding a total derivative of the form 
$\tn_\mu\xi^\mu$ with
\begin{equation}
 \xi^\mu=\frac{\BBB}{2\Lambda^2\sqrt{\fac}}\left[\tn^\mu\phi\Box\phi-\tn^\mu\tn^\nu\phi\partial_\nu\phi\right].
\end{equation}
This simplifies the action to
\onecolumngrid
\begin{equation} 
\frac{\mathcal{L}_g}{\sqrt{-\tg}}=\sqrt{\fac}\frac{\tilde{R}}{2}+\frac{\BBB}{2\Lambda^2\sqrt{\fac}}\left[
(\Box\phi)^2-\tn_\mu\tn_\nu\phi\tn^\mu\tn^\nu\phi\right]-\frac{B(\phi)B_\phi(\phi)}{\Lambda^2\sqrt{\fac}}\left[
2X\Box\phi+\tn^\mu\phi\tn_\mu\tn_\nu\phi\tn^\nu\phi\right ].
\end{equation}
\twocolumngrid
One can see that the quartic terms are in Horndeski form but there is still one cubic term that does not fit. This too can be removed by subtracting 
a second total derivative $\tn_\mu\zeta^\mu$ with
\begin{equation}
 \zeta^\mu=\sqrt{\fac}\frac{B_\phi(\phi)}{B(\phi)}\tn^\mu\phi.
\end{equation}
The action then becomes
\onecolumngrid
\begin{align}
\frac{\mathcal{L}_g}{\sqrt{-\tg}}&=\sqrt{\fac}\frac{\tilde{R}}{2}+\frac{\BBB}{2\Lambda^2\sqrt{\fac}}\left[
(\Box\phi)^2-\tn_\mu\tn_\nu\phi\tn^\mu\tn^\nu\phi\right] \nonumber-\frac{B_\phi(\phi)}{B(\phi)\Lambda^2\sqrt{\fac}}\left[1+\frac{4\BBB 
X}{\Lambda^2}\right]\Box\phi\\&+2X\left[\frac{B_{\phi\phi}}{\BBB}\sqrt{\fac}-\frac{B_\phi^2(\phi)}{\BBB\sqrt{\fac}}\right],
\end{align}
\twocolumngrid
which is now in the Horndeski form.

Next, we need to transform $\mathcal{L}_\phi$. This is a lot simpler since one only needs to transform the metric determinant and the metric 
appearing in the kinetic term using (\ref{eq:tg2}) to find
\begin{equation}
\frac{\mathcal{L}_\phi}{\sqrt{-\tg}}=\sqrt{\fac}\left[X-\frac{2\BBB X^2}{\Lambda^4\left(\fac\right)}-V(\phi)\right]. 
\end{equation}
The action is then in Horndeski form with 
\onecolumngrid
\begin{align}
G_4(\phi,X)&=\frac{\mpl^2}{2}\sqrt{\fac}\label{eq:horn4}\\  
G_3(\phi,X)&=-\mpl^2\frac{B_\phi(\phi)}{B(\phi)\Lambda^2\sqrt{\fac}}\left[1+\frac{4\BBB 
X}{\Lambda^2}\right]\label{eq:horn3}\\
G_2(\phi,X)&=\mpl^2\sqrt{\fac}\left[\frac{2XB_{\phi\phi}}{\BBB}-\frac{2XB_\phi^2(\phi)}{\BBB\left({\fac}\right)}+X-\frac{2\BBB 
X^2}{\Lambda^4\left(\fac\right)}-V(\phi)\right].\label{eq:horn2}
\end{align}
\twocolumngrid
Note that a similar action was obtained by \cite{RenauxPetel:2011uk}.

\subsection{The Field Equations}\label{sec:field_eqns}

Given equations (\ref{eq:horn4})--(\ref{eq:horn2}), it is clear that the resulting field equations will be cumbersome and complicated. Since we are 
only interested in the homogeneous and isotropic Friedmann equations it is simplest to first reduce the action to minisuperspace using the coordinates
\begin{equation}
 \dd \tilde{s}^2=-N^2(t)\dd t^2+a(t)^2\dd\vec{x}^2;\quad \phi=\phi(t).
\end{equation}
The Friedmann and Klein-Gordon equations can then be found using the Euler-Lagrange equations for $N(t)$, $a(t)$ and $\phi(t)$ and setting $N(t)=1$. 
Setting $B(\phi)=e^{\beta\phi}$\footnote{We do this for simplicity, it is not necessary to specialise at this stage but leaving the function general 
results in a far longer expression.} one finds 
\onecolumngrid
\begin{align}
 S[N(t),a(t),\phi(t)]&=\int\dd 
t\frac{a(t)^3}{\sqrt{\facc}}\left[3\gamma\frac{\dot{a}^2}{a^2N^2}-3\gamma\frac{\dot{a}\dot{N}}{aN^2}+3\gamma\beta\frac{\dot{a} \dot { \phi } } { aN } 
-\beta\gamma\frac { \dot { N } \dot { \phi } }{N^2 }+\frac{\dot{\phi}^2}{2}\right.\nonumber\\&\left. 
-V(\phi)N^2\left(\facc\right)+\beta^2\frac{e^{2\beta\phi}\dot{\phi}^4}{\Lambda^2N^3}+3\frac{\ddot{a}}{aN}\left(\facc\right)+\beta\gamma\frac{\ddot{
\phi } } { N } +3\frac{e^{2\beta\phi}\dot{a}\dot{\phi}\ddot{\phi}}{\Lambda^2aN^3}\right ],
\end{align}
\twocolumngrid
where
\begin{equation}
 \gamma=\faccc.
\end{equation}
The Euler-Lagrange equation for $N(t)$ yields the Friedmann equation \eqref{3} (after setting $N=1$), which can be used in the Euler-Lagrange 
equations for $a(t)$ and $\phi(t)$ to find equations \eqref{4} and \eqref{5}.

\section{Dynamical System when $\beta=\lambda/2$}\label{app:reddyn}

In this Appendix we present the dynamical system after substituting the constraint \eqref{eq:cons} into \eqref{eq1}--\eqref{eq3} to eliminate $Y$. 
They are:
\onecolumngrid
\begin{align}
U_1\frac{\dd X}{\dd N}&=-X \left(3 X^4 \left(4 \left(\mu ^2-1\right) Z^{\frac{4}{3}}+3 Z^{8/3}+1-4 \mu ^4\right)+2 \sqrt{6} \lambda  
X^3 Z 
\left(2 \mu ^2-2 \left(2 \mu ^2+1\right) Z^{\frac{4}{3}}+Z^{8/3}+1\right)\right.\nonumber\\&\left.-6 X^2 Z^2 \left(Z^{\frac{4}{3}}-1\right) \left(2 
\mu ^2+4 
Z^{\frac{4}{3}}-3\right)+\sqrt{6} \lambda  X Z^3 \left(Z^{\frac{4}{3}}-1\right)^2-3 Z^4 \left(Z^{\frac{4}{3}}-1\right)^2\right),\\
U_2\frac{\dd Y}{\dd N}&=\mu  X \left(3 X^4 \left(2 \mu ^2+Z^{\frac{4}{3}}-1\right) \left(2 \mu ^2-3 Z^{\frac{4}{3}}+1\right)+2 \sqrt{6} \lambda  
\left(2 \mu 
^2+1\right) X^3 Z \left(Z^{\frac{4}{3}}-1\right)\nonumber\right.\\&\left.+6 X^2 Z^2 \left(Z^{\frac{4}{3}}-1\right) \left(2 \mu ^2+2 
Z^{\frac{4}{3}}-3\right)-\sqrt{6} \lambda 
 X Z^3 \left(Z^{\frac{4}{3}}-1\right)^2+3 Z^4 \left(Z^{\frac{4}{3}}-1\right)^2\right),\\
 \frac{\dd Z}{\dd N}&=\frac{3 X^2 \left(\sqrt{6} \lambda  X Z^{\frac{4}{3}}-6 Z^{7/3}+6 Z-\sqrt{6} \lambda  \left(2 \mu ^2+1\right) X\right)}{2 
\left(X^2 
\left(-2 \mu ^2+3 Z^{\frac{4}{3}}-1\right)-Z^{10/3}+Z^2\right)}
\end{align}
where
\begin{align}
 U_1&=2 Z^2 \left(Z^{\frac{4}{3}}-1\right) \left(X^2 \left(2 \mu ^2-3 Z^{\frac{4}{3}}+1\right)+Z^2 
\left(Z^{\frac{4}{3}}-1\right)\right)\\
U_2&=Z^2 \sqrt{2-2 Z^{\frac{4}{3}}} \left(Z^{\frac{4}{3}}-1\right) \left(X^2 \left(2 \mu ^2-3 
Z^{\frac{4}{3}}+1\right)+Z^2 \left(Z^{\frac{4}{3}}-1\right)\right).
\end{align}
\twocolumngrid
Note that only two of these are independent since differentiating the constraint one has
\begin{equation}
 2\frac{X'}{X}=2\frac{Y'}{Y}+\frac{4Z^{\frac{1}{3}}}{3(1-Z^{\frac{4}{3}})}\frac{Z'}{Z}.
\end{equation}
It is straightforward to verify that the dynamical system above indeed satisfies this relation.

\end{document}